\documentclass[12pt]{iopart}
\expandafter\let\csname equation*\endcsname\relax
\expandafter\let\csname endequation*\endcsname\relax
\usepackage{amsmath}  
\usepackage{amsfonts}
\usepackage{amssymb}  
\usepackage{graphicx} 
\usepackage{enumerate}
\usepackage[hidelinks]{hyperref}
\begin{document}
\newcommand{\ds}{\displaystyle}
\newcommand{\be}{\begin{equation}}
\newcommand{\en}{\end{equation}}
\newcommand{\bea}{\begin{eqnarray}}
\newcommand{\ena}{\end{eqnarray}}

\title{Casimir--like wormholes}
\author{Mauricio Cataldo}
\ead{mcataldo@ubiobio.cl} 
\address{Departamento de
F\'\i sica, Universidad del Bío-B\'io, Casilla 5-C,
Concepci\'on, Chile.\\}
\address{Centro de Ciencias Exactas, Universidad del Bío-Bío, Casilla 447, Chillán, Chile. \\}

\begin{abstract}
We investigate spherically symmetric spacetimes sourced by matter with an energy density scaling as $\rho \propto 1/r^{4}$ and obeying the equation of state $p_r = \omega \rho$. We derive the most general solution of Einstein’s field equations for this class of sources and identify a broad family of geometries, including black holes, naked singularities, and traversable wormholes, parametrized by the integration constants and the equation of state parameter $\omega$. We determine the conditions for traversability and show that only two classes of matter can support traversable wormholes: Casimir--type matter with $\omega > 0$ and phantom matter with $\omega < -1$, whereas quintessence and conventional dark energy are excluded by the flare--out condition at the throat. For the Casimir value $\omega = 3$, we find that the transverse pressure approaches $p_t \to -\rho$ near the central naked singularity at $r = 0$, implying that the complete algebraic structure of the Casimir stress--energy tensor, characterized by $\rho \propto -1/r^{4}$, $p_r = 3\rho$, and $p_t = -\rho$, is dynamically recovered in the vicinity of the origin. This behavior is not imposed \textit{a priori} but emerges directly from Einstein’s field equations. Adopting a purely classical framework in which the amplitude of the energy density is treated as a free parameter, we obtain traversable wormhole solutions with throat radii ranging from microscopic to astrophysical scales. These Casimir--like configurations share the functional structure of the Casimir stress--energy tensor, $\rho\propto r^{-4}$ and, for $\omega=3$, $p_r=3\rho$, but are not generated by the physical Casimir effect. Within this setting, Garattini’s Planck--scale Casimir wormhole is recovered as a particular case when the quantum Casimir energy scale is imposed. We further show that the total amount of exotic matter required is finite for arbitrary $\omega$ and scales linearly with the throat radius. We analyze the conditions for human traversability, including the acceleration felt by the traveler, tidal forces and transit times, and show that a radial free fall satisfies all of them simultaneously for throat radii $r_0\gtrsim10^{8}$~m. Finally, we present embedding diagrams illustrating three distinct geometric regimes depending on the value of $\omega$.


\end{abstract}

\submitto{\CQG}

\maketitle 


\section{Introduction}
In the study of traversable wormholes, two main approaches are commonly employed in the literature. The first, and perhaps most prevalent, is the geometric approach, in which one postulates specific forms for the metric functions, namely, the shape function $b(r)$ and the redshift function $\phi(r)$, and subsequently derives the energy density and pressures required to support such a geometry through the Einstein field equations. This approach, while straightforward, often leads to matter sources whose physical interpretation remains unclear. The second approach proceeds in the opposite direction: one specifies the matter content through equations of state relating the pressures to the energy density, and then solves the Einstein field equations to determine the metric functions, while simultaneously imposing the conditions necessary for traversability, such as the throat condition, the flaring--out requirement, and the absence of horizons. 

In the present work, we adopt this second approach, although the traversability conditions are imposed only after the general solution has been obtained. Among the various candidates for exotic matter capable of supporting a traversable wormhole, the Casimir energy stands out as the only known source of negative energy density that has been experimentally verified in the laboratory. The Casimir effect, first predicted theoretically by Casimir in 1948~\cite{Casimir1948} and experimentally confirmed in subsequent years~\cite{Sparnaay1957,Sparnaay1958,Lamoreaux1997,Bressi2002}, constitutes one of the most striking manifestations of quantum vacuum fluctuations~\cite{Milonni1994}. This phenomenon arises between two plane, parallel, closely spaced, uncharged metallic plates placed in vacuum, where the modification of the electromagnetic zero--point energy gives rise to an attractive force. The renormalized energy density associated with the Casimir effect is given by
\begin{equation}
\rho_C = -\frac{\hbar c \pi^2}{720\, a^4},
\end{equation}
where $a$ denotes the separation between the parallel plates. The negative sign of the energy density, together with the associated radial pressure $p_r = 3\rho_C$, defines an equation of state of the form $p_r = \omega \rho$ with $\omega = 3$. The corresponding stress--energy tensor is
\begin{equation}
T_{\mu\nu}
=
\frac{\hbar c \pi^{2}}{720\, a^{4}}
\,\mathrm{diag}(-1,-3,1,1)\,,
\end{equation}
from which it follows that the tangential pressure satisfies $p_t = -\rho$. This stress--energy tensor is both traceless and divergenceless, reflecting the conformal invariance of the underlying massless field and ensuring local energy--momentum conservation.

The existence of negative energy densities is of central importance in the context of General Relativity, as it implies the violation of the null energy condition (NEC). This violation is precisely the requirement for the existence of traversable wormholes, which are solutions of Einstein's field equations that connect two asymptotically flat regions of spacetime through a throat~\cite{Morris1988,Visser1995}. 

Given the quantum nature of the Casimir effect, it has been suggested as a natural candidate to source traversable wormholes~\cite{Visser1995,Morris1988WH}. In this context, the classical Einstein field equations must be replaced by their semiclassical counterpart
\begin{equation}
G_{\mu\nu} = \kappa \langle T_{\mu\nu} \rangle_{\text{ren}},
\end{equation}
where $\kappa = 8\pi G/c^4$ and $\langle T_{\mu\nu} \rangle_{\text{ren}}$ represents the renormalized stress--energy tensor of the quantum fields.

The possibility of using Casimir energy to support a traversable wormhole was first examined by Morris, Thorne and Yurtsever~\cite{Morris1988WH}, who concluded that the plate separation should be smaller than the electron Compton wavelength, rendering the physical apparatus unrealizable. Subsequently, Visser~\cite{Visser1995} proposed a model where the mass of the plates compensates for the negative energy density, forbidding the creation of a traversable wormhole.

More recently, Garattini~\cite{Garattini2019} revisited this problem by promoting the constant plate separation $a$ to a radial coordinate $r$, effectively considering a Casimir energy density of the form $\rho(r) \propto -1/r^4<0$. By imposing the equation of state $p_r = \omega \rho$, and solving the Einstein field equations, he found, for a generic value of $\omega$, a solution describing a traversable wormhole of Planckian size.

Wormholes supported by matter obeying a prescribed equation of state have been widely studied. The linear relation $p_r=\omega\rho$ with constant $\omega<-1$ defines the phantom-energy wormholes of Refs.~\cite{LoboPhantom2005,Sushkov2005}. These constructions were subsequently extended to the generalized Chaplygin gas~\cite{Lobo2006}, to polytropic phantom energy~\cite{Jamil2010} and quintessence polytropic energy~\cite{Manna}, to evolving geometries supported by polytropic matter~\cite{Cataldo2013}, to variable equation-of-state parameters~\cite{Cataldo2017,Koussour,Parsaei}, including finite-size configurations supported by anisotropic dark energy~\cite{Cataldo2017} and extensions to $f(Q,T)$ gravity~\cite{Koussour}, and to polynomial equations of state combined with power-law shape functions~\cite{Parsaei2020}. Power-law density profiles $\rho\propto r^{-n}$ satisfy the flare--out condition for $n>2$~\cite{Kim2013}. Following Garattini's work, Casimir wormholes have been examined in $2+1$ dimensions, where the Casimir energy density and tension alone cannot sustain them~\cite{Alencar2021}, in $D$ dimensions~\cite{Oliveira2022}, with the Yang--Mills Casimir energy~\cite{Santos2023}, and with generalized uncertainty principle corrections~\cite{Jusufi2020,Muniz2024,Battista2024}, while traversable wormholes supported by quantum effects of charged massless fermions have been constructed in four dimensions~\cite{Maldacena2023}. The present work differs from these studies in two main respects. First, rather than prescribing the shape function or imposing the throat condition from the outset, we obtain the most general solution of the field equations for $\rho\propto r^{-4}$ and $p_r=\omega\rho$, with arbitrary $\omega$ and arbitrary sign of the amplitude. This family contains black holes, naked singularities and traversable wormholes, and reveals a Casimir--type structure near the origin. Second, by treating the amplitude as a free classical parameter, we obtain macroscopic wormholes and a phantom branch, both absent in the quantum Casimir case.

In the present work, we extend the study of Garattini~\cite{Garattini2019} by considering wormhole configurations sustained by exotic matter satisfying an equation of state of the form $p_r(r) = \omega \rho(r)$, for which the energy density scales as $\rho(r) \propto \pm \, 1/r^{4}$. Of particular interest is the case $\omega = 3$, which corresponds to the Casimir equation of state. While Garattini's analysis focused on the quantum Casimir effect, which necessarily leads to wormholes of Planckian size, we adopt a broader perspective by treating the Casimir--type equation of state from a purely classical standpoint. This approach allows us to explore traversable wormhole solutions of arbitrary size, from Planckian to macroscopic and even cosmological scales, without restricting ourselves to the specific energy scales dictated by quantum vacuum fluctuations. Furthermore, our analysis encompasses both negative and positive energy densities, revealing that the mathematical structure of the field equations admits traversable solutions in both regimes. 

We emphasize that our aim is not to address the mechanism by which such wormholes might grow to macroscopic sizes, but rather to demonstrate that the space of solutions is considerably richer than previously recognized,  showing that Einstein's field equations admit Casimir--like wormhole solutions at scales far beyond the Planck length.

Throughout this work, we distinguish between Casimir wormholes, namely solutions in which the energy density $\rho(r) = A/(\kappa r^{4})$ has an amplitude $A<0$ fixed by the quantum Casimir energy scale, necessarily yielding Planckian throat radii, and Casimir--like wormholes, which retain the same functional form $\rho(r) = A/(\kappa r^{4})$ and the barotropic equation of state $p_r = \omega \rho$, while allowing the parameter $A$ to take arbitrary positive or negative classical values, thereby permitting throat radii of unrestricted size.

Casimir--like wormholes should therefore not be regarded as generated by the physical Casimir effect, whose quantum energy scale is retained only in the Planckian case. What they share with the Casimir effect is the functional structure of its stress--energy tensor, namely an energy density scaling as $\rho\propto r^{-4}$ and, for the Casimir value $\omega=3$, the same ratio $p_r/\rho=3$. In particular, the power law $\rho\propto r^{-4}$ is a phenomenological choice, inherited from Garattini's promotion of the plate separation to the radial coordinate, and it is not singled out by the wormhole geometry itself, since any profile $\rho\propto r^{-n}$ with $n>2$ satisfies the flare--out condition~\cite{Kim2013}. We adopt $n=4$ for two reasons. First, $n=4$ reproduces the scaling $\rho_{_C}\propto a^{-4}$ of the Casimir energy density, the only negative energy density measured in the laboratory, and is therefore the natural choice for extending Garattini's quantum construction to classical scales. Second, it is the case in which the field equations can be solved exactly in closed form. As shown in section~\ref{sec:general_solution}, the causal structure of the general solution is then governed by the quadratic polynomial $r^2-c_1r+A$, whereas for integer $n>4$ the corresponding polynomial is of degree $n-2$, and the complete classification of the solutions obtained here would no longer be available in transparent form.

To keep the terminology unambiguous, we reserve the expressions \emph{Casimir effect}, \emph{physical Casimir effect} and \emph{quantum Casimir effect} for the actual quantum phenomenon between conducting plates, and the terms \emph{Casimir wormholes} and \emph{Casimir--like wormholes} for the two classes of geometries defined above. The adjective \emph{Casimir--type} refers instead to the matter content. We call the barotropic relation $p_r=\omega\rho$, together with $\rho\propto r^{-4}$, a \emph{Casimir--type equation of state}, since it generalizes the Casimir relation $p_r=3\rho$, and we call \emph{Casimir--type matter} the case of negative energy density with $\omega>0$, for which $\rho$ and $p_r$ have the same signs as in the Casimir tensor. The \emph{Casimir value} $\omega=3$ reproduces, in addition, the Casimir ratio $p_r/\rho$.

Throughout this paper, we use units in which $c=1$ and $\kappa = 8 \pi G$.

The paper is organized as follows. In section~\ref{sec:general}, we review the Morris--Thorne wormhole geometry, present the Einstein field equations for an anisotropic matter source, and introduce the Casimir--type equation of state. In section~\ref{sec:general_solution}, we derive the most general spherically symmetric solution with an energy density scaling as $\rho \propto r^{-4}$, explicitly determining the metric functions and discussing the physical significance of the central naked singularity, where the complete algebraic structure of the Casimir stress--energy tensor is dynamically recovered. In section~\ref{sec:traversable}, we focus on traversable wormhole configurations, establish the corresponding traversability conditions, and classify the types of matter capable of supporting such geometries. We recover Garattini's Planck--scale Casimir wormhole as a particular case, explore macroscopic Casimir--like wormholes, compute the total amount of exotic matter required, analyze the relevant traversability constraints, and present embedding diagrams for different values of the equation of state parameter. Finally, in section~\ref{sec:conclusions}, we summarize our results and discuss their implications.

\section{Traversable wormhole geometry and field equations}
\label{sec:general}

\subsection{The Morris--Thorne metric}\label{subsec:MT}

We consider a static, spherically symmetric line element of the form
\begin{equation}
ds^2 = -e^{2\phi(r)} dt^2 + \frac{dr^2}{1-b(r)/r} + r^2 d\Omega^2,
\label{eq:metric}
\end{equation}
where $\phi(r)$ and $b(r)$ are arbitrary functions of the radial coordinate $r$. This metric provides a general framework for describing static, spherically symmetric spacetimes and encompasses a wide variety of gravitational configurations, including wormhole geometries. Here $d\Omega^2 = d\theta^2 + \sin^2\theta\, d\varphi^2$ denotes the line element of the unit two--sphere, and $(t,r,\theta,\varphi)$ are standard Schwarzschild--like coordinates, with $t \in (-\infty,+\infty)$, $r \in (0,+\infty)$, $\theta \in [0,\pi]$, and $\varphi \in [0,2\pi)$.

In the context of wormhole physics, the metric~\eqref{eq:metric} is commonly referred to as the Morris--Thorne line element~\cite{Morris1988}. The function $\phi(r)$ is known as the redshift function, as it governs the gravitational redshift of signals propagating through the spacetime, while $b(r)$ is the shape function, which encodes the spatial geometry of the wormhole and determines the shape of equatorial slices in embedding diagrams.

For the metric \eqref{eq:metric} to describe a traversable wormhole, several conditions must be satisfied:

\begin{enumerate}
    \item  Throat condition: At the throat $r = r_0$, we require $b(r_0) = r_0$.
    \item  Flaring--out condition: The throat must flare outward, which requires $(b - b'r)/b^2 > 0$ at the throat. Combined with the throat condition, this implies $b'(r_0) < 1$.    
    \item  Asymptotic flatness: For the spacetime to be asymptotically flat, we require $b(r)/r \to 0$ and $\phi(r) \to 0$ as $r \to \infty$.    
    \item  Absence of horizons: To ensure traversability, there must be no event horizons, which means $e^{2\phi(r)}$ must be finite and non--zero for all $r \geq r_0$. This requires $\phi(r)$ to be finite everywhere.    
    \item  Proper radial distance: The proper radial distance 
\begin{equation}
\ell(r) = \pm \int_{r_0}^{r} \frac{dr'}{\sqrt{1-b(r')/r'}} \label{eq:proper_distance}
\end{equation}    
must be finite throughout the spacetime.
\end{enumerate}

\subsection{Einstein field equations}
In this work, we restrict our attention to \emph{classical} wormholes, namely spacetimes that solve the classical Einstein field equations. 
Although the equation of state adopted for the matter source is motivated by quantum considerations, specifically by the functional structure of the Casimir effect, both the gravitational field and the matter content are treated entirely within a classical framework. We do not compute quantum expectation values of the stress--energy tensor, nor do we address semiclassical backreaction effects. Instead, we prescribe an energy density and pressure profile whose functional form is motivated by the Casimir stress--energy tensor.

This effective description allows us to investigate, in a controlled and fully classical setting, the geometric implications of exotic matter obeying the equation of state $p_r=\omega\rho$, without restricting the analysis to Planckian regimes or invoking the machinery of quantum field theory in curved spacetime. 
In this sense, the Casimir effect serves as a physically motivated guide for the choice of matter content, while the resulting wormhole geometries remain solutions of the classical Einstein equations.

The geometry of spacetime is thus governed by the Einstein field equations $G_{\mu\nu} = \kappa T_{\mu\nu}$, where $G_{\mu\nu} = R_{\mu\nu} - \frac{1}{2}g_{\mu\nu}R$ is the Einstein tensor, $T_{\mu\nu}$ is the stress--energy tensor of the matter content, and $\kappa$ is the Einstein gravitational constant defined as $\kappa = 8\pi G/c^4$, with $G$ being Newton's gravitational constant and $c$ the speed of light in vacuum. In geometrized units where $G = c = 1$, we simply have $\kappa = 8\pi$.

\subsection{Anisotropic matter source}

Given the spherical symmetry of the spacetime, the most general form of the stress--energy tensor compatible with the metric \eqref{eq:metric} is that of an anisotropic fluid. In an orthonormal reference frame, this tensor takes the diagonal form
\begin{equation}
T_{\hat{\mu}\hat{\nu}} = \text{diag}\left(-\rho(r), p_r(r), p_t(r), p_t(r)\right),
\label{eq:SET}
\end{equation}
where $\rho(r)$ is the energy density, $p_r(r)$ is the radial pressure measured in the direction perpendicular to the $t = \text{const}$ hypersurfaces, and $p_t(r)$ is the tangential (or transverse) pressure measured in the angular directions. The anisotropy of the fluid is characterized by the difference $\Delta = p_t - p_r$, which in general is non--vanishing for wormhole spacetimes.

\subsection{Components of the field equations}
The Einstein field equation for the metric \eqref{eq:metric} are given by:
\begin{align}
\kappa \rho &= \frac{b'}{r^2}, \label{eq:field_tt}\\[2mm]
\kappa p_r &= \frac{2r(r-b)\phi' - b}{r^3}, \label{eq:field_rr}\\[2mm]
\kappa p_t &= \frac{1}{r^3}\left\{r^2(r-b)\phi'' + \left[r(r-b)\phi' \right.\right. \nonumber\\
&\quad \left.\left. - \frac{b'r - b}{2}\right](\phi' r + 1)\right\}, \label{eq:field_thth}
\end{align}
where primes denote derivatives with respect to $r$ and we have omitted the explicit $r$-dependence for brevity.

Additionally, the conservation of the stress--energy tensor, $\nabla_\mu T^{\mu\nu} = 0$, provides the hydrostatic equilibrium equation
\begin{equation}
p_r'(r) = \frac{2}{r}\left(p_t(r) - p_r(r)\right) - \left(\rho(r) + p_r(r)\right)\phi'(r),
\label{eq:conservation}
\end{equation}
which is not independent but can be derived from the Bianchi identities. This equation describes how the pressure gradients balance the gravitational and anisotropic forces within the wormhole.

\subsection{Energy conditions}

Wormholes are supported by exotic matter sources that typically violate all classical energy conditions~\cite{Visser1995}. These conditions are physically motivated constraints imposed on the energy--momentum tensor, intended to capture fundamental properties expected of the matter content in a given spacetime. For anisotropic matter sustaining a traversable wormhole, the energy--momentum  tensor takes the form $T^{\mu}_{\nu}=\mathrm{diag}(-\rho,p_r,p_t,p_t)$. The energy conditions are formulated in terms of the following inequalities~\cite{Visser1995}:
\begin{enumerate}
\item Weak Energy Condition (WEC): \\ $\rho \geq 0$, $\rho + p_r \geq 0$ and $\rho + p_t \geq 0$.
\item Null Energy Condition (NEC): \\ $\rho + p_r \geq 0$ and $\rho + p_t \geq 0$.
\item Dominant Energy Condition (DEC): \\ $\rho \geq |p_r|$ and $\rho \geq |p_t|$, or equivalently $\rho + p_r \geq 0$, $\rho + p_t \geq 0$ and $\rho- p_r \geq 0$ and $\rho - p_t \geq 0$.
\item Strong Energy Condition (SEC):\\ $\rho + p_r + 2p_t \geq 0$.
\end{enumerate}

At the wormhole throat, the flare--out condition combined with the Einstein equations for energy density $\rho$ and radial pressure $p_r$ implies
\begin{equation}
\frac{1 - b'(r_0)}{r_0^2} > 0 \quad \Rightarrow \quad \rho(r_0) + p_r(r_0) < 0,
\end{equation}
which demonstrates that the NEC must be violated at the throat.

\subsection{Equation of state}\label{subsec:eos}
To close the system of Einstein field equations associated with the metric~\eqref{eq:metric}, we adopt a barotropic equation of state relating the radial pressure to the energy density,
\begin{equation}
p_r(r) = \omega \rho(r),
\label{eq:eos}
\end{equation}
where $\omega$ is the equation of state parameter. For matter obeying this relation, the violation of the NEC at the throat $r=r_0$ requires
\begin{equation}
(1+\omega)\,\rho(r_0) < 0.
\end{equation}

This condition plays a central role in assessing the viability of traversable wormhole solutions. In particular, for Casimir--type matter characterized by $\omega = 3$ and a negative energy density $\rho<0$, the inequality above is automatically satisfied, confirming that Casimir energy provides a natural source for NEC--violating matter capable of sustaining traversable wormholes.

As discussed earlier, the Casimir effect between parallel conducting plates is characterized by the specific value $\omega = 3$. Under this choice, the corresponding stress--energy tensor assumes the form
\begin{equation}
T_{\mu\nu}^{\text{Casimir}} = \frac{A}{r^4}\,
\mathrm{diag}\!\left(-1, -3, 1, 1\right),
\label{eq:SET_casimir}
\end{equation}
where the constant $A$ sets the amplitude of the energy density. This explicit form makes manifest the anisotropic pressure structure intrinsic to the Casimir vacuum and serves as the prototype for the Casimir--type sources considered throughout this work.

However, in our analysis we shall keep $\omega$ as a free parameter. 
This generalization serves two purposes: first, it allows us to explore how the wormhole properties associated with a quartic energy density, $\rho \propto r^{-4}$, depend on the equation of state; second, it encompasses other physically motivated scenarios in which the radial pressure and energy density remain proportional but with a different proportionality constant.

It is worth remarking on the unusual nature of the Casimir equation of state parameter $\omega = 3$. In the context of classical relativistic fluids, an ideal fluid is typically characterized by $\omega < 1$, which ensures that the speed of sound $c_s = \sqrt{dp/d\rho} = \sqrt{\omega}\,c$ remains subluminal. The limiting case $\omega = 1$, known as stiff matter, corresponds to a fluid where the speed of sound equals the speed of light. For $\omega > 1$, the classical interpretation breaks down as it would imply superluminal sound propagation if $dp/d\rho$ were interpreted as a sound speed. The Casimir value $\omega = 3$ lies well within this exotic regime, in full accordance with its quantum origin and the associated violation of the classical energy conditions.

We stress that none of this casts doubt on the physical consistency of the static configuration itself, since it satisfies Einstein's field equations exactly, with $\rho(r)$ and $p_r(r)$ sourcing a genuine curved geometry whose physical consequences are derived throughout this paper, including the tidal forces, the embedding structure, and the exotic matter content. What is not determined by this static background solution alone is how the same matter would respond to time-dependent perturbations. Interpreting the slope $dp_r/d\rho=\omega$ of the equation of state $p_r=\omega\rho$ as the squared speed of small perturbations requires an additional assumption, namely that matter displaced from equilibrium responds according to the same relation $p_r(\rho)$ that links pressure and density across the static configuration, as would be the case, for instance, if the matter were barotropic or isentropic. This is a hypothesis about the dynamics of the matter, not a consequence of the static solution, and it is not guaranteed to hold for the matter sourcing this geometry. Indeed, the barotropic relation $p_r=\omega\rho$ relates pressure and density across different points of a single equilibrium configuration; it does not, by itself, specify an equation of motion for the matter, from which an actual propagation speed for perturbations could be derived. For instance, for an ordinary minimally coupled scalar field the instantaneous slope $\dot p/\dot\rho$ need not coincide with the causal sound speed governing its perturbations; during slow-roll inflation $\dot p/\dot\rho\simeq-1$, whereas $c_s^2=1$ exactly~\cite{Ellis2007}. More generally, de la Macorra and Vucetich~\cite{deLaMacorra2004} demonstrate this point by explicit construction, comparing distinct microscopic realizations of a scalar-field source, namely a canonical field, a tachyonic field ($m^2<0$), a phantom field (reversed-sign kinetic term), and $k$-essence models with different choices of the kinetic functions $K(\phi)$, $F(X)$. For the canonical, tachyonic and phantom cases they study, the actual signal propagation speed never exceeds $c$, regardless of how far the formal slope $dp/d\rho$ departs from unity or even diverges; for $k$-essence, the propagation speed is set by the specific kinetic functions chosen, and remains below $c$.
In short, a given equation of state $p_r=\omega\rho$ does not by itself fix how perturbations propagate. A barotropic perfect fluid obeying it would have $c_s^2=\omega$, whereas a canonical scalar field has $c_s^2=1$ whatever its equation of state~\cite{Ellis2007}, so the propagation speed is set by the dynamics of the matter and not by the equation of state alone.

In the present work we do not specify which matter fills the wormhole. The relation $p_r=\omega\rho$ is introduced as an algebraic closure condition on the components of an effective, static stress-energy tensor, and not as the equation of motion of an identified field or fluid. Our analysis therefore does not rely on a Lagrangian, a microscopic field content, or any other dynamical realization of the source. As a result, the actual propagation speed of perturbations cannot be determined within the present framework. 

Consistently with this, the actual matter content needed to produce the quantum Casimir effect itself is the electromagnetic field in its vacuum state, whose own propagating excitations, photons, travel at the speed of light, illustrating, in the one case where the microscopic realization is in fact known, that the formal value $\omega=3$ is compatible with luminal propagation.

By contrast, since the microscopic realization of the macroscopic source considered here is left unspecified, the formal value $dp_r/d\rho=\omega=3$ cannot, by itself, settle whether perturbations of the source propagate causally. Likewise, the foundational constructions of phantom-energy wormholes~\cite{LoboPhantom2005,Sushkov2005}, with $p_r=\omega\rho$ and constant $\omega<-1$, proceed directly from the energy density and pressure profile to the Einstein field equations without identifying an underlying field or fluid. A similar situation arises in a recent study of Casimir wormholes with an isotropic effective source~\cite{Muniz2024}, where, for the Casimir value $\omega=3$, the squared sound speed obtained from the equation of state exceeds unity over the whole range of radii considered, a feature that is reported without affecting the subsequent analysis of the solution.

It is important to emphasize that, in general, the transverse pressure $p_t(r)$ is \emph{not} specified independently, but is instead fully determined by the Einstein field equations once the energy density $\rho(r)$, the radial pressure $p_r(r)$, and the spacetime geometry encoded in the functions $b(r)$ and $\phi(r)$ are fixed. This feature represents a key distinction between the present approach and several previous treatments in the literature.

As will be shown, the resulting transverse pressure $p_t(r)$ generally departs from the simple relation $p_t=-\rho$ that characterizes the Casimir effect.

\section{Spherically symmetric solution with $\rho(r) \propto r^{-4}$}\label{sec:general_solution}

Although the metric~\eqref{eq:metric} introduced by Morris and Thorne was originally formulated to describe traversable wormhole geometries, its structure is considerably more general and provides a convenient parametrization for any static, spherically symmetric spacetime, regardless of whether a wormhole interpretation is ultimately realized. Accordingly, in this section we employ the Morris--Thorne form as a general framework to derive the most general spherically symmetric solution of the Einstein field equations for a spacetime filled with a Casimir--type energy density scaling as $\rho \propto r^{-4}$, systematically obtaining the metric functions $b(r)$ and $\phi(r)$ without imposing any a priori restrictions beyond spherical symmetry. This approach reveals the full landscape of admissible solutions, from which specific traversable wormhole configurations can be identified by imposing appropriate boundary conditions.

\subsection{Energy density and equation of state}

We consider an anisotropic fluid characterized by an energy density of the form
\begin{equation}
\rho(r) = \frac{A}{\kappa r^4},
\label{eq:rho_general}
\end{equation}
where $A$ has dimensions of length squared and determines both the magnitude and sign of the energy density. For $A < 0$, the energy density is negative, corresponding to the exotic matter required to sustain traversable wormholes. Conversely, $A > 0$ yields positive energy density configurations.

The radial pressure is related to the energy density through a barotropic equation of state
\begin{equation}
p_r(r) = \omega \rho(r) = \frac{\omega A}{\kappa r^4},
\label{eq:pr_general}
\end{equation}
where $\omega$ is the equation of state parameter. The value $\omega = 3$ corresponds to the Casimir equation of state, although our analysis remains valid for arbitrary $\omega$.

\subsection{Determination of the shape function}

From Eq.~\eqref{eq:field_tt}, combined with the energy density \eqref{eq:rho_general}, we obtain the differential equation $b'(r) = \frac{A}{r^2}$. Direct integration yields the shape function
\begin{equation}
b(r) = c_1-\frac{A}{r},
\label{eq:b_integrated}
\end{equation}
where $c_1$ is an integration constant to be determined by boundary conditions. The structure of this solution already reveals an important feature: the shape function naturally accommodates both wormhole geometries (when appropriate conditions on $c_1$ and $A$ are imposed) and other spherically symmetric spacetimes.

\subsection{Determination of the redshift function}

The redshift function $\phi(r)$ is determined by the radial component of the Einstein equations. Substituting the shape function \eqref{eq:b_integrated} into Eq.~\eqref{eq:field_rr} and imposing the equation of state \eqref{eq:pr_general}, we obtain
\begin{equation}
\frac{2}{\kappa r}\left(1 - \frac{c_1}{r} + \frac{A}{r^2}\right)\phi'(r) - \frac{1}{\kappa r^3}\left(c_1 - \frac{A}{r}\right) = \frac{\omega A}{\kappa r^4}.
\label{eq:phi_eq}
\end{equation}

This first--order linear differential equation can be integrated analytically, yielding
\begin{eqnarray}
\phi(r) &=& \frac{1-\omega}{4}\ln(r^2 - c_1 r + A) + \frac{\omega - 1}{2}\ln r \nonumber\\
&& + \frac{c_1(\omega + 1)}{2\sqrt{c_1^2 - 4A}}\text{arctanh}\left(\frac{2r - c_1}{\sqrt{c_1^2 - 4A}}\right) + c_2, \,\,\,\,\,\,\,\,\,\,\,\,\,\,\,
\label{eq:phi_solution}
\end{eqnarray}
where $c_2$ is a second integration constant, typically fixed by asymptotic flatness requirements.

The solution \eqref{eq:phi_solution} can be cast into a more transparent exponential form. Introducing the roots of the quadratic polynomial $P(r) = r^2 - c_1 r + A$,
\begin{equation}
r_\pm = \frac{c_1 \pm \sqrt{c_1^2 - 4A}}{2},
\label{eq:roots_def}
\end{equation}
the metric function $g_{tt} = -e^{2\phi(r)}$ takes the form
\begin{equation}
e^{2\phi(r)} = e^{2c_2} \, r^{\omega-1} \left(r - r_+\right)^{\alpha_+} \left(r - r_-\right)^{\alpha_-},
\label{eq:exp2phi_general}
\end{equation}
where the exponents are given by
\begin{align}
\alpha_+ &= \frac{1-\omega}{2} + \frac{c_1(\omega+1)}{2\sqrt{c_1^2 - 4A}}, \label{eq:alpha_minus}\\[2mm]
\alpha_- &= \frac{1-\omega}{2} - \frac{c_1(\omega+1)}{2\sqrt{c_1^2 - 4A}}. \label{eq:alpha_plus}
\end{align}

This factorized representation is particularly useful for analyzing the causal structure of the spacetime. The explicit dependence of $e^{2\phi(r)}$ on the factors $(r - r_+)$ and $(r - r_-)$ makes the location and nature of special surfaces in the geometry immediately transparent, thereby revealing the rich causal structure of this family of solutions. The roots $r_\pm$ determine where the temporal component of the metric may vanish or diverge, while the exponents $\alpha_\pm$ govern the precise behavior at these surfaces. When an exponent is positive, the corresponding factor vanishes at the root, potentially signaling a horizon; when negative, the factor diverges, indicating a singularity; and when zero, the factor becomes unity, rendering the root a regular point of the metric. This interplay between the positions $r_\pm$ and the exponents $\alpha_\pm$ gives rise to a diverse landscape of spacetime geometries, which includes black holes, naked singularities, and regular configurations, all emerging from a single family of solutions parametrized by $c_1$, $A$, and $\omega$. The particular geometry realized depends on the specific values of these parameters, which determine whether the roots lie in the physical domain $r > 0$ and whether the corresponding exponents lead to regular or singular behavior.

\subsection{Transverse pressure}\label{subsec:pt}

Unlike the energy density and radial pressure, which are specified as input through the equation of state, the transverse pressure $p_t(r)$ emerges as an output determined by the $(\hat{\theta}\hat{\theta})$ component of the Einstein equations. This is a distinctive feature of anisotropic fluids in general relativity: once two components of the stress--energy tensor are prescribed, the remaining component is fixed by the field equations and the Bianchi identities.

By substituting the shape function \eqref{eq:b_integrated} and the redshift function \eqref{eq:exp2phi_general} into Eq.~\eqref{eq:field_thth}, the transverse pressure can be expressed entirely in terms of the physical parameters $c_1$, $A$, and $\omega$:
\begin{equation}
\begin{split}
\kappa\, p_{t}(r) &= \frac{A \left(-4 \omega\, r^{2}+c_{1} \left(5 \omega +1\right) r +A \left(\omega^{2}-4 \omega -1\right)\right)}{4 r^{4} \left(r^2 - c_1 r + A\right)}\\[2mm]
&= \frac{A \left(-4 \omega\, r^{2}+c_{1} \left(5 \omega +1\right) r +A \left(\omega^{2}-4 \omega -1\right)\right)}{4 r^{4} \left(r -r_{\texttt{+}} \right) \left(r -r_{\texttt{-}} \right)}.
\end{split}
\label{eq:pt_general}
\end{equation}
It is instructive to examine the limiting case $A = 0$. In this limit, Eqs.~\eqref{eq:b_integrated} and \eqref{eq:exp2phi_general} reduce to $b(r) = c_1$ and $e^{2\phi} \propto (r - c_1)/r$, respectively, recovering the Schwarzschild solution, as expected. This consistency check confirms that our general solution correctly encompasses the vacuum case and demonstrates the singular nature of the $r = 0$ region in the absence of exotic matter.

Another relevant aspect concerns the behavior of the transverse pressure near the origin. 
In the limit $r \to 0$, the transverse pressure behaves as
\begin{equation}
\kappa p_{t}(r) \longrightarrow \frac{A\left(\omega^{2} - 4\omega - 1\right)}{4\,r^{4}}. \label{eq:C}
\end{equation}
Remarkably, for $\omega = 3$ this expression reduces to
\begin{equation}
\kappa p_t(r) \longrightarrow -\frac{A}{r^4}. \label{eq:Comega}
\end{equation}
Recalling that $\kappa \rho(r) = A/r^4$ and $\kappa p_r(r) = 3A/r^4$, this immediately yields the relation
\[
p_t = -\rho.
\]
This is precisely the equation of state that characterizes the original Casimir effect in flat spacetime, where the stress--energy tensor takes the form $T_{\mu\nu} \propto \mathrm{diag}(-1, -3, 1, 1)$. Therefore, sufficiently close to the singularity at $r = 0$, the complete Casimir equation of state is recovered for all pressure components, namely $p_r = 3\rho$ and $p_t = -\rho$. This demonstrates that, in the near--origin regime, the spherically symmetric solution with $\rho \propto r^{-4}$ exactly reproduces the stress--energy tensor structure of the original Casimir effect. 

This result is particularly noteworthy, as it is the Einstein field equations themselves that enforce this specific matter behavior in the neighborhood of $r = 0$. It is worth emphasizing that the relation $p_t = -\rho$ is not imposed as an input condition, but rather emerges dynamically as a prediction of the theory. Furthermore, this feature appears to have gone unnoticed in previous analyses of Casimir wormholes~\cite{Garattini2019}.

To assess where this recovery is physically meaningful, it is useful to expand the exact solution near the origin. For $\omega=3$, Eqs.~\eqref{eq:rho_general} and~\eqref{eq:pt_general} give
\begin{equation}
\frac{p_t}{\rho}=-1+\frac{3c_1}{A}\,r+\mathcal{O}\!\left(\frac{r^2}{A},\frac{c_1^2r^2}{A^2}\right),
\label{eq:pt_expansion}
\end{equation}
so that the Casimir--type structure is already approximately attained for radii small compared with the scales of the solution, $r\ll\sqrt{|A|}$ and $r\ll|A|/|c_1|$, and not only at $r=0$. On the other hand, the curvature, of order $|A|/r^4$, reaches the Planck scale only for $r\lesssim(|A|\,\ell_P^2)^{1/4}$, where $\ell_P$ is the Planck length, and it is only there that classical general relativity is expected to fail. For a macroscopic amplitude, $|A|\gg\ell_P^2$, there is therefore an intermediate window,
\begin{equation}
(|A|\,\ell_P^2)^{1/4}\ll r\ll\sqrt{|A|},
\label{eq:window}
\end{equation}
together with $r\ll|A|/|c_1|$, in which the classical description remains reliable and the Casimir--type structure is already recovered to good approximation. For instance, for $\sqrt{|A|}=1$~m the window extends from about $4\times10^{-18}$~m to about $1$~m. This is an order--of--magnitude estimate, which assumes that classical general relativity breaks down at Planckian curvature, and the lower bound could be higher if the effective description of the source fails before. For the Planckian Casimir wormhole, with $|A|\sim\ell_P^2$, the two scales coincide and no such window exists. We also stress that this discussion concerns the general solution of this section, since the traversable wormholes of section~\ref{sec:traversable} are defined for $r\ge r_0$ and do not contain the origin.

This behavior contrasts sharply with the asymptotic region $r \rightarrow \infty$, where all components of the stress--energy tensor vanish, ensuring asymptotic flatness. The solution thus interpolates between a regime near the origin in which the matter acquires the complete Casimir--type stress--energy structure and a vacuum--like behavior at spatial infinity.

In summary, the general solution obtained in this section, characterized by the shape function \eqref{eq:b_integrated} and the redshift function given in Eq.~\eqref{eq:exp2phi_general}, admits multiple geometric interpretations depending on the boundary conditions imposed on the integration constants $c_1$ and $c_2$. Different choices of these constants select distinct members from the family of solutions, leading to spacetimes with fundamentally different causal structures, including black holes, naked singularities, or regular configurations. 

In the following section, we shall focus on the subset of solutions describing traversable wormholes. This is achieved by imposing the throat condition $b(r_0) = r_0$, which fixes the integration constant $c_1$ in terms of the throat radius $r_0$ and the parameter $A$, thereby selecting from the general family those spacetimes possessing a wormhole throat at $r = r_0$. This procedure necessarily introduces deviations from the complete Casimir--type structure of the matter in the vicinity of $r = 0$.

\subsection{On the physical relevance of central naked singularities}
\label{subsec:naked_singularity}

Although in the next section we will focus on the analysis of traversable wormhole configurations, in this subsection we wish to emphasize the fact that the spherically symmetric solutions presented in this work may exhibit a curvature singularity at the origin $r=0$, where the Casimir--type energy density $\rho = A/(\kappa r^4)$ diverges, as well as the radial pressure $p_r=3 \rho$ and the transverse pressure $p_t=-\rho$. The existence of a strong curvature singularity is confirmed by the divergence of the Kretschmann scalar as $r \to 0$. When this singularity is not covered by an event horizon, it constitutes a naked singularity. A natural question arises regarding the physical viability of such configurations.

The existence of naked singularities at the center of spherically symmetric spacetimes has been extensively studied in the context of gravitational collapse and static equilibrium configurations. Joshi, Malafarina, and Narayan~\cite{Joshi2011,Joshi2014} demonstrated that a slowly evolving gravitationally collapsing matter cloud can asymptotically settle into a static, spherically symmetric equilibrium configuration with a naked singularity at the center. In their models, the energy density profile diverges as $\rho_e \propto R^{-n}$ with $n > 0$ near the origin, and the existence of a strong curvature singularity is confirmed by the divergence of the Kretschmann scalar at $r = 0$. Crucially, the absence of trapped surfaces ensures that the singularity is not covered by an event horizon. Although the matter content in Refs.~\cite{Joshi2011,Joshi2014} (perfect fluid or tangential pressure) differs from the Casimir--type source considered here, the essential geometric feature, a central naked singularity arising from a divergent energy density profile, is common to both scenarios. This comparison is, however, only a structural analogy. In Refs.~\cite{Joshi2011,Joshi2014} the naked singularity arises as the end state of a specific collapse process with a different matter content, whereas no formation mechanism is proposed here, so that their results cannot be taken as a physical justification of the singular branch obtained in this work.

A key insight from Refs.~\cite{Joshi2011,Joshi2014} is the physical interpretation of such singularities: they represent regions where the energy density reaches arbitrarily large values, signaling the breakdown of classical general relativity and the expected onset of quantum gravitational effects.

The presence of a central naked singularity is therefore not, by itself, inconsistent within classical general relativity, but its physical status remains open. As discussed in section~\ref{subsec:pt}, the classical solution remains reliable down to radii at which the curvature approaches the Planck scale, see Eq.~\eqref{eq:window}, whereas arbitrarily close to $r=0$ the classical description is expected to fail. This singular region does not affect the traversable wormholes studied in the next section, which do not contain the origin. Whether quantum gravitational effects ultimately resolve the central singularity, possibly replacing it with a Planck scale core or connecting the geometry to another spacetime region, remains an open problem that lies beyond the scope of the present classical treatment.

\section{Traversable wormholes with $\rho(r) \propto r^{-4}$}\label{sec:traversable}
In this section, we consider traversable wormhole configurations supported by an energy density of Casimir type, $\rho \propto r^{-4}$. In the input expression~\eqref{eq:rho_general}, the sign of the constant $A$ determines whether the energy density is negative ($A < 0$), as in the genuine quantum Casimir effect, or positive ($A > 0$). As established in the Introduction, the distinction between Casimir wormholes and Casimir--like wormholes lies not in the sign of $A$, but in whether the amplitude is fixed by the quantum Casimir energy scale or treated as a free classical parameter.

\subsection{Determination of the shape function and the flare--out condition}
In order to obtain a wormhole solution, the throat condition $b(r_0) = r_0$ must be imposed on the shape function~\eqref{eq:b_integrated}. This requirement fixes the integration constant as
\begin{equation}
c_1 = \frac{r_0^2 + A}{r_0}.
\label{eq:c1_value}
\end{equation}
Substituting this result back into Eq.~\eqref{eq:b_integrated}, the shape function becomes
\begin{equation}
b(r)=r_0 + \frac{A(r - r_0)}{r\, r_0},
\label{eq:b_final}
\end{equation}
which explicitly satisfies the throat condition at $r = r_0$.

The flare--out condition at the throat requires that $b'(r_0) < 1$. Using Eqs.~\eqref{eq:field_tt} and~\eqref{eq:rho_general}, we obtain
\begin{equation}
b'(r) = \frac{A}{r^2},
\end{equation}
and consequently
\begin{equation}
b'(r_0) = \frac{A}{r_0^2} < 1 \quad \Rightarrow \quad A < r_0^2.
\label{eq:flaring}
\end{equation}

For Casimir--like wormholes with $A < 0$, this condition is automatically satisfied for any throat radius $r_0 > 0$. In contrast, for Casimir--like configurations with $A > 0$, the flare--out condition imposes an upper bound on the parameter $A$ in terms of the throat size.

\subsection{Rewriting the general solution}
Substituting Eq.~\eqref{eq:c1_value} into the expressions for the roots~\eqref{eq:roots_def}, we find $c_1^2 - 4A = \frac{(r_0^2 - A)^2}{r_0^2}$. It then follows immediately that
\begin{equation}
\sqrt{c_1^2 - 4A} = \frac{|r_0^2 - A|}{r_0}.
\end{equation}

This result yields two possible branches:
\begin{eqnarray*}
r_0^2 - A > 0 \quad \text{(i.e., } A < r_0^2\text{)},
\\
A - r_0^2 > 0 \quad \text{(i.e., } A > r_0^2\text{)}.
\end{eqnarray*}
The second branch must be excluded, since it violates the flare--out condition~\eqref{eq:flaring}. 
Therefore, only the branch satisfying $A < r_0^2$ is physically admissible for traversable wormholes and will be considered in what follows.

We further note that the constraint $r_0^{2} - A > 0$ is automatically satisfied for all configurations with $A < 0$, corresponding to negative energy density, as well as for those with $0 < A < r_0^{2}$, which describe positive energy densities below the critical value. In this regime the roots simplify to
\begin{align}
r_+ &= \frac{1}{2}\!\left(\frac{r_0^2 + A}{r_0} - \frac{r_0^2 - A}{r_0}\right)
     = \frac{A}{r_0}, \label{eq:rplus_simplified} \\[2mm]
r_- &= \frac{1}{2}\!\left(\frac{r_0^2 + A}{r_0} + \frac{r_0^2 - A}{r_0}\right)
     = r_0. \label{eq:rminus_simplified}
\end{align}

With the simplified roots~\eqref{eq:rplus_simplified} and~\eqref{eq:rminus_simplified}, 
the general solution~\eqref{eq:exp2phi_general} reduces to
\begin{equation}
e^{2\phi(r)} = e^{2c_2}\, r^{\omega-1}
\left(r - \frac{A}{r_0}\right)^{\alpha_+}
\left(r - r_0\right)^{\alpha_-},
\label{eq:exp2phi_simplified}
\end{equation}
where the exponents become, after substitution,
\begin{align}
\alpha_+ &= \frac{\omega r_0^2 + A}{A - r_0^2}, 
\label{eq:alpha_plus_final} \\[2mm]
\alpha_- &= \frac{-\omega A - r_0^2}{A - r_0^2}.
\label{eq:alpha_minus_final}
\end{align}

For configurations satisfying the flare--out condition $A < r_0^2$, the factor
$\left(r - \dfrac{A}{r_0}\right)^{\alpha_+}$ remains strictly positive throughout the physical domain $r \ge r_0$, and therefore does not obstruct traversability. 
In contrast, the factor $(r - r_0)^{\alpha_-}$ vanishes at the throat, so the sign and magnitude of the exponent $\alpha_-$ must be chosen with care in order to avoid the formation of horizons or curvature singularities at $r = r_0$.

\subsection{Conditions for traversability and energy conditions}

The redshift function \eqref{eq:exp2phi_simplified} describes a traversable wormhole only if  $e^{2\phi(r)}$ is finite and non--zero for all $r \geq r_0$. Therefore, the factor $(r - r_0)^{\alpha_-}$ is the critical term:

\begin{itemize}
\item If $\alpha_- > 0$: $e^{2\phi} \to 0$ as $r \to r_0$, producing a horizon (black hole).
\item If $\alpha_- < 0$, one finds that $e^{2\phi} \to \infty$ as $r \to r_0$, indicating a curvature singularity. Unlike the central singularity at $r = 0$, this one occurs at a finite radius $r = r_0$. Furthermore, the absence of an event horizon in this spacetime renders the singularity naked.
\item If $\alpha_- = 0$: $e^{2\phi}$ remains finite and non--zero at the throat, yielding a traversable wormhole.
\end{itemize}

Therefore, the traversability condition reads
\begin{equation}
\alpha_- = 0  
\quad \Rightarrow \quad 
A = -\frac{r_0^2}{\omega}.
\label{eq:traversability_alpha_plus}
\end{equation} 
The same traversability condition was obtained by Garattini~\cite{Garattini2019} for the Casimir amplitude, in the form $\omega r_1^2=r_0^2$ with $r_1^2=\pi^3\ell_P^2/90$, the remaining cases yielding either a horizon or a singularity at $r=r_0$.

Accordingly, the wormhole geometry is completely specified by the shape function~\eqref{eq:b_final} and the redshift function~\eqref{eq:exp2phi_simplified}, which assume the explicit forms
\begin{align}
b(r) &= r_{0} - \frac{r_{0}\left(r - r_{0}\right)}{\omega\, r}, \label{eq:bwh} \\
\mathrm{e}^{2\phi(r)} &=
\mathrm{e}^{2 c_{\texttt{2}}}
\left(\frac{\omega\, r}{\omega\, r + r_{0}}\right)^{\omega - 1}. \label{eq:ephi}
\end{align}
It is important to emphasize that the shape function~\eqref{eq:bwh} is well defined only for $\omega \neq 0$, and therefore the $\omega = 0$ branch does not belong to this class of solutions. To obtain the corresponding solution for $\omega = 0$, one must instead consider Eq.~\eqref{eq:phi_eq}. A direct inspection of this case reveals that an event horizon inevitably forms at $r = r_{0}$, rendering the resulting geometry non--traversable. Consequently, this branch is excluded from the present analysis.

For the solutions described by Eqs.~\eqref{eq:bwh} and~\eqref{eq:ephi}, the associated matter components are given by
\begin{eqnarray}
\kappa\,\rho(r) &=& -\frac{r_{0}^{2}}{\omega\, r^{4}}, \label{eq:rhoCWH} \\
\kappa\,p_{r}(r) &=& -\frac{r_{0}^{2}}{r^{4}}, \label{eq:prCWH} \\
\kappa\,p_{t}(r) &=&
\frac{r_{0}^{2}\!\left[(4r - r_{0})\omega^{2} + 4\omega r_{0} +r_{0}\right]}
{4\omega\left(\omega r + r_{0}\right) r^{4}}. \label{eq:ptCWH}
\end{eqnarray}
Equations~\eqref{eq:bwh}--\eqref{eq:ptCWH} coincide with the traversable solution obtained by Garattini~\cite{Garattini2019} for a generic $\omega$. In that case, however, the amplitude is fixed by the quantum Casimir energy scale, so that the traversability condition ties the throat radius to the Planck length. Here the same geometry holds for an arbitrary amplitude $A$, and the throat radius is a free parameter.

It is interesting to note that the radial pressure is always negative, regardless of the value of the equation of state parameter $\omega$. This behavior is expected for traversable wormhole configurations, which require exotic matter violating the null energy condition. It is worth noting that the radial pressure~\eqref{eq:prCWH} is independent of $\omega$. This follows directly from combining the equation of state with the traversability condition. Since $p_r=\omega\rho$ and, from Eq.~\eqref{eq:traversability_alpha_plus}, $A=-r_0^2/\omega$, Eq.~\eqref{eq:rhoCWH} gives
\begin{equation}
\kappa p_r(r) = \omega\,\kappa\rho(r) = \omega\left(-\frac{r_0^2}{\omega\, r^4}\right) = -\frac{r_0^2}{r^4},
\end{equation}
so that the factor of $\omega$ introduced by the equation of state is exactly cancelled by the factor of $1/\omega$ already present in the amplitude fixed by traversability. This cancellation is not a numerical coincidence but a direct algebraic consequence of imposing both conditions simultaneously, and holds for every value of $\omega$ permitted by Table~\ref{tab:matter_types}.

On the other hand, the sign of the energy density depends on $\omega$: it is negative for $\omega > 0$ and positive for $\omega < 0$. However, a positive energy density does not automatically imply 
that the corresponding configuration represents a wormhole supported by quintessence or dark energy; the viability of such solutions must be assessed by examining both the flare--out condition and the energy conditions at the throat.

To determine traversability, one must examine the flare--out condition. 
From Eq.~\eqref{eq:bwh}, the first derivative of the shape function is given by
\begin{equation}
b'(r) = -\frac{r_0^2}{\omega r^2},
\end{equation}  
which, evaluated at the throat, yields
\begin{equation}
b'(r_0) = -\frac{1}{\omega}.
\end{equation}  
For the flare--out condition to be satisfied, we require $b'(r_0) < 1$, that is,
\begin{equation}
-\frac{1}{\omega} < 1.  \label{eq:FOC}
\end{equation}
This condition excludes configurations supported by quintessence ($-1/3 < \omega < 0$) and conventional dark energy ($-1 < \omega < -1/3$), as these yield $b'(r_0) > 1$. Consequently, only two classes of matter satisfy the flare--out condition: phantom matter with $\omega < -1$, 
and Casimir--type matter with $\omega > 0$. 

As we show now, the above conclusion is further corroborated by the analysis of the energy conditions. Using Eqs.~\eqref{eq:rhoCWH}--\eqref{eq:ptCWH}, we can examine the 
energy conditions in detail. 

From Eq.~\eqref{eq:rhoCWH} it follows that the sign of the energy density depends on the value of $\omega$: for $\omega < 0$ one has $\rho > 0$, corresponding to a positive energy density, whereas for $\omega > 0$ the energy density becomes negative, $\rho < 0$. On the other hand, combining Eqs.~\eqref{eq:rhoCWH} and~\eqref{eq:prCWH}, we obtain
\begin{equation}
\kappa \rho(r) + \kappa p_r(r) = -\frac{r_0^2(1 + \omega)}{\omega r^4}, 
\label{eq:AFOC}
\end{equation}
which, evaluated at the throat, gives
\begin{equation}
\kappa \rho(r_0) + \kappa p_r(r_0) = -\frac{1 + \omega}{\omega r_0^2}.
\end{equation}
The flare--out condition~\eqref{eq:FOC} implies that the combination $\rho(r_0) + p_r(r_0)$ is necessarily negative at the throat. In particular, for $\omega < -1$ one finds $\rho(r) > 0$ together with $\kappa\,[\rho(r_0) + p_r(r_0)] < 0$, whereas for $\omega > 0$ it follows that $\rho(r) < 0$ and $\kappa\,[\rho(r_0) + p_r(r_0)] < -1/r_0^2 < 0$. Consequently, in both cases the WEC, NEC and DEC are violated at the throat, as required for the existence of a traversable wormhole.

Furthermore, Eq.~\eqref{eq:AFOC} shows that, for $\omega < -1$ and $\omega > 0$, the quantity $\rho(r) + p_r(r)$ remains negative for all $r \geq r_0$. As a result, the weak, null, and dominant energy conditions are violated throughout the entire spacetime, and not merely at the throat.

The remaining energy conditions, evaluated at the throat, are given by
\begin{eqnarray}
\kappa\rho(r_0) + \kappa p_t(r_0) &=& \frac{3(\omega - 1)}{4\omega r_0^2}, \\
\kappa \rho(r_0) + \kappa p_r(r_0) + 2\kappa p_t(r_0) &=&
\frac{\omega - 1}{2\omega r_0^2}.
\end{eqnarray}
From these expressions, it follows that both $\rho(r_0) + p_t(r_0)$ and $\rho(r_0) + p_r(r_0) + 2p_t(r_0)$ are positive for $\omega < -1$ and $\omega > 1$, while they become negative in the interval $0 < \omega < 1$. Accordingly, the tangential pressure satisfies the weak, null, and dominant energy conditions for $\omega < -1$ and $\omega > 1$. In this regime, the strong energy condition is also fulfilled.

Table~\ref{tab:matter_types} summarizes the ranges of the equation of state parameter $\omega$ for which Casimir--like wormholes are traversable.

\begin{table}[h]
\caption{Classification of matter types according to the equation of state parameter $\omega$ for spacetimes with $\rho \propto r^{-4}$. The table shows whether each matter type satisfies the NEC and the flare--out condition required for traversability. Only Casimir--type matter ($\omega > 0$) and phantom matter ($\omega < -1$) can support traversable wormholes; both classes yield Casimir--like wormholes when the amplitude $A$ is treated as a free parameter. Quintessence and conventional dark energy are excluded because they fail to satisfy the flare--out condition. The case $\omega = -1$ yields $b'(r_0) = 1$, which does not satisfy the strict inequality $b'(r_0) < 1$ required for traversability.}
\label{tab:matter_types}
\begin{indented}
\item[]\begin{tabular}{lcccc}
\mr
\textbf{Matter type} & \textbf{$\omega$} & \textbf{NEC} & 
\textbf{Flare--out} & \textbf{Trav.} \\
\mr
Casimir--type & $\omega > 0$ & Violated & Satisfied & Yes \\
\mr
Quintessence & $-\frac{1}{3} < \omega < 0$ & Satisfied & Violated & No \\
\mr
Dark energy & $-1 < \omega < -\frac{1}{3}$ & Satisfied & Violated & No \\
\mr
Cosm.\ const.\ & $\omega = -1$ & Saturated & Not satisfied & No \\
\mr
Phantom & $\omega < -1$ & Violated & Satisfied & Yes \\
\mr
\end{tabular}
\end{indented}

\end{table}

We now comment on the naked singularity located at a finite radius $r = r_0$. Since the flare--out condition requires $A < r_0^2$, it follows from Eq.~\eqref{eq:alpha_minus_final} that, whenever $\omega A < r_0^2$, one has $\alpha_- < 0$. In this regime, the tangential pressure takes the form
$\kappa\, p_{t}(r) =
-\frac{\bigl[r (5\omega + 1)(r_0^2 + A) + \bigl(-4\omega r^2 + A(\omega^2 - 4\omega - 1)\bigr) r_0\bigr] A}
{4 (r - r_0)(A - r r_0)\, r^{4}}$. Taking the limit $r_0 \to 0$, we find
$\kappa\, p_{t}(r) \longrightarrow -\frac{A(5\omega + 1)}{4\, r^{4}}$,
which, for the Casimir value $\omega = 3$, reduces to $\kappa\, p_{t}(r) \longrightarrow -\frac{4A}{r^{4}}$. Therefore, in this case, when the sphere $r = r_0$ shrinks to zero radius, the spherically symmetric solution with $\rho \propto r^{-4}$ fails to reproduce the stress--energy tensor structure of the original Casimir effect. This behavior contrasts sharply with that of the central naked singularity at $r = 0$, where the complete algebraic structure of the Casimir stress--energy tensor is exactly recovered, as discussed in connection with Eqs.~\eqref{eq:C} and~\eqref{eq:Comega}.

\subsection{The Traversable Planckian Casimir Wormhole}
\label{sec:casimir}

Among the family of traversable wormhole solutions obtained above, a case of particular physical significance arises when the matter source corresponds to the genuine quantum Casimir effect. This connection, first established by Garattini~\cite{Garattini2019}, emerges naturally when we identify the parameter $A$ with the Casimir energy scale.

The key observation is that our general energy density~\eqref{eq:rho_general}, $\kappa \rho(r) = \frac{A}{r^4}$, becomes identical to the quantum Casimir energy density when we set
\begin{equation}
A = -\frac{\kappa \hbar c \pi^2}{720} = -\frac{\pi^3 \ell_P^2}{90},
\label{eq:A_casimir}
\end{equation}
where $\ell_P = \sqrt{\hbar G/c^3}$ is the Planck length. This identification, which lies at the heart of Garattini's approach, ties the amplitude of the energy density to fundamental constants and thereby constrains the geometric parameters of the resulting wormhole.

For the Casimir equation of state $\omega = 3$, the traversability condition~\eqref{eq:traversability_alpha_plus} requires $A = -r_0^2/3$. Combining this with the quantum identification~\eqref{eq:A_casimir}, we obtain
\begin{equation}
\frac{r_0^2}{3} = \frac{\pi^3 \ell_P^2}{90},
\end{equation}
which fixes the throat radius at
\begin{equation}
r_0 = \sqrt{\frac{\pi^3}{30}} \, \ell_P \approx 1.02 \, \ell_P.
\end{equation}
This result confirms that wormholes sourced by the quantum Casimir effect with $\omega = 3$ are necessarily of Planckian size, in agreement with Garattini's original finding~\cite{Garattini2019}.

A similar analysis applies to the case $\omega = 1$, which corresponds to the Ellis--Bronnikov wormhole~\cite{Ellis1973,Bronnikov1973}. In this case, the traversability condition yields $A = -r_0^2$, and the quantum identification~\eqref{eq:A_casimir} leads to
\begin{equation}
r_0^2 = \frac{\pi^3 \ell_P^2}{90},
\end{equation}
giving a throat radius
\begin{equation}
r_0 = \sqrt{\frac{\pi^3}{90}} \, \ell_P \approx 0.59 \, \ell_P,
\end{equation}
which is sub--Planckian. Thus, both the Casimir wormhole ($\omega = 3$) and the Ellis--Bronnikov wormhole ($\omega = 1$), when sourced by quantum vacuum fluctuations, possess throats of order the Planck length, far too small to be traversable in any practical sense.

It is important to emphasize, however, that the Planckian size is not an intrinsic feature of the wormhole geometry itself, but rather a consequence of identifying the amplitude $A$ with the specific energy scale of the quantum Casimir effect. From a purely classical standpoint, the parameter $A$ in our general solution~\eqref{eq:rho_general} remains unconstrained, and the throat radius $r_0$ can assume any positive value. This observation motivates the broader class of Casimir--like wormholes, which we discuss in the following subsection: solutions that retain the equation of state $p_r = \omega \rho$, with $\kappa \rho \propto r^{-4}$, characteristic of the Casimir effect, while treating the amplitude $A$ as a free parameter, thereby allowing for macroscopic and potentially astrophysical throat sizes.

\subsection{Traversable Casimir--like wormholes}
\label{sec:casimir-like}

As established in the previous subsection, the Planckian size of Casimir wormholes arises solely from identifying the parameter $A$ with the quantum vacuum energy scale~\eqref{eq:A_casimir}. However, from a purely classical perspective, there is no fundamental reason to impose this restriction. In this subsection, we explore the broader class of Casimir--like wormholes, defined as traversable wormhole solutions satisfying the equation of state $p_r = \omega \rho$ with $\kappa \rho \propto r^{-4}$, but with the amplitude $A$ treated as a free parameter unconstrained by quantum considerations.

This approach aligns with the standard methodology employed in modern wormhole physics, where one postulates the existence of exotic matter satisfying certain equations of state and investigates the resulting geometric and traversability properties, leaving aside the question of its microscopic origin. In this spirit, we treat the Casimir--type equation of state as a phenomenological ansatz that may arise from yet unknown physics at macroscopic scales.

For the Casimir value $\omega = 3$, the traversability condition~\eqref{eq:traversability_alpha_plus} requires
\begin{equation}
A = -\frac{r_0^2}{3},
\label{eq:A_macroscopic}
\end{equation}
where $r_0$ is now a free parameter representing the throat radius. The corresponding energy density and pressures are given by
\begin{eqnarray}
\kappa \rho(r) &=& -\frac{r_0^2}{3r^4}, \label{eq:rho_macro}\\
\kappa p_r(r) &=& -\frac{r_0^2}{r^4}, \label{eq:pr_macro}\\
\kappa p_t(r) &=& \frac{r_0^2(9r + r_0)}{3r^4(3r + r_0)}, \label{eq:pt_macro}
\end{eqnarray}
while the metric functions read
\begin{eqnarray}
b(r) &=& \frac{2r_0}{3} + \frac{r_0^2}{3r}, \label{eq:b_macro}\\
e^{2\phi(r)} &=& \left(\frac{3r}{3r + r_0}\right)^2, \label{eq:phi_macro}
\end{eqnarray}
where we have set $e^{2c_2} = 1$ to ensure asymptotic flatness.

The throat radius can range from microscopic to macroscopic and even astrophysical values, with the amplitude of the energy density scaling accordingly through Eq.~\eqref{eq:A_macroscopic}. This freedom opens the possibility of studying Casimir--like wormholes as classical objects of arbitrary size, amenable to the same geometric and traversability analyses typically performed for other wormhole models in the literature.

To illustrate this point, we present in Table~\ref{tab:scales} the energy density at the throat, $|\rho(r_0)| = 1/(3\kappa r_0^2)$ (that is, $c^4/(24\pi G r_0^2)$ in SI units), together with its mass equivalent $|\rho(r_0)|/c^2$, for various choices of throat radius spanning from Planckian to astrophysical scales, and compare them with several reference densities.

\begin{table}[h]
\caption{Energy density at the throat for wormholes with Casimir--type matter ($\omega = 3$) at different scales. The Planckian case corresponds to the Casimir wormhole when the amplitude is fixed by the quantum vacuum energy scale~\eqref{eq:A_casimir}; the remaining entries represent Casimir--like wormholes with $A$ treated as a free parameter. The required energy densities decrease as $r_0^{-2}$ with increasing throat size. The third column gives the energy density and the fourth its mass equivalent. The row $r_0=5.5\times10^{7}$~m corresponds to the minimum throat radius allowed by the radial tidal constraint of section~\ref{sec:tidal}. The last three rows give, for comparison, the energy density of the laboratory Casimir effect for a plate separation $a=1\,\mu$m, the nuclear saturation density, and a representative neutron--star core density. Since the wormhole energy densities are negative, this comparison only indicates orders of magnitude.} 
\label{tab:scales}
\begin{indented}
\item[]{\setlength{\tabcolsep}{4pt}\begin{tabular}{lccc}
\mr
\textbf{Scale} & \textbf{$r_0$} & \textbf{$|\rho(r_0)|$   (J/m$^3$)} & \textbf{$|\rho(r_0)|/c^2$ (kg/m$^3$)} \\
\mr
Planckian & $10^{-35}$ m & $\sim 10^{112}$ & $\sim 10^{95}$ \\
\mr
Nuclear & $10^{-15}$ m & $\sim 10^{72}$ & $\sim 10^{55}$ \\
\mr
Macroscopic & $1$ m & $\sim 10^{42}$ & $\sim 10^{25}$ \\
\mr
Stellar & $10^{3}$ m & $\sim 10^{36}$ & $\sim 10^{19}$ \\
\mr
Astrophysical & $10^{6}$ m & $\sim 10^{30}$ & $\sim 10^{13}$ \\
\mr
Tidal bound (section~\ref{sec:tidal}) & $5.5\times10^{7}$ m & $\sim 5\times10^{26}$ & $\sim 6\times10^{9}$ \\
\mr
\multicolumn{4}{l}{\textit{Reference values}} \\
\mr
Laboratory Casimir ($a=1\,\mu$m) & -- & $\sim 4\times10^{-4}$ & $\sim 5\times10^{-21}$ \\
\mr
Nuclear saturation & -- & $\sim 2\times10^{34}$ & $\sim 3\times10^{17}$ \\
\mr
Neutron--star core & -- & $\sim 10^{35}$ & $\sim 10^{18}$ \\
\mr
\end{tabular}}
\end{indented}

\end{table}

A similar analysis holds for the Ellis--Bronnikov case $\omega = 1$, where the traversability condition yields $A = -r_0^2$. In this case, the energy density at the throat is $|\rho(r_0)| = 1/(\kappa r_0^2)$, which exhibits the same scaling behavior with throat size.

We emphasize that our aim is not to address the mechanism by which such exotic matter might be produced or how wormholes might grow to macroscopic sizes, but rather to demonstrate that the mathematical structure of Einstein's equations coupled to Casimir--type matter admits a rich family of traversable configurations extending far beyond the Planck length. The physical realizability of these solutions remains an open question, intimately connected to the broader problem of exotic matter sources in general relativity.

The values in Table~\ref{tab:scales} make explicit the distinction between the mathematical existence of macroscopic Casimir--like wormholes and their physical realizability. In terms of mass--equivalent density, the requirement decreases as $r_0^{-2}$, falls below the nuclear saturation density for $r_0\gtrsim 8$~km, and reaches $\sim 6\times10^{9}$~kg/m$^3$ for the throat radius $r_0\approx5.5\times10^{7}$~m that satisfies the radial tidal constraint derived in section~\ref{sec:tidal}. This comparison is, however, only indicative, since the required energy density is negative. The relevant benchmark is the laboratory Casimir effect, whose energy density is about thirty orders of magnitude smaller than the value required at $r_0\approx5.5\times10^{7}$~m, and even further below the values required for smaller throats. Moreover, while the local density decreases with the throat size, the total amount of exotic matter grows linearly with $r_0$ (see Table~\ref{tab:exotic_matter}), reaching about $2.5\times10^{4}$ solar masses in the same case. No known form of matter, nor any experimentally accessible negative--energy configuration, can supply these requirements. The solutions obtained here therefore establish what the field equations permit, not what is physically realizable.

\subsection{Total amount of exotic matter}
\label{sec:exotic_matter}

A crucial aspect in the study of traversable wormholes is the quantification of the total amount of exotic matter required to sustain the geometry. This is typically assessed through the volume integral quantifier, which measures the total violation of the averaged null energy condition (ANEC) throughout the spacetime~\cite{Visser2003,Lobo2005}. This quantity is defined as
\begin{equation}
I_V = \int [\rho(r) + p_r(r)] \, dV,
\label{eq:IV_def}
\end{equation}
where the integration extends over the entire wormhole spacetime. For the Morris--Thorne metric~\eqref{eq:metric}, using the volume element $dV = 4\pi r^2 dr$, this becomes
\begin{equation}
I_V = 4\pi \int_{r_0}^{\infty} [\rho(r) + p_r(r)] \, r^2 \, dr.
\label{eq:IV_integral}
\end{equation}

Substituting Eq.~\eqref{eq:AFOC} into Eq.~\eqref{eq:IV_integral} we obtain
\begin{eqnarray}
\kappa I_V &=& -\frac{4\pi r_0^2(1 + \omega)}{\omega} \int_{r_0}^{\infty} \frac{dr}{r^2},
\end{eqnarray}
and performing the integration we find
\begin{equation}
I_V = -\frac{4\pi r_0(1 + \omega)}{\kappa \omega}.
\label{eq:IV_result}
\end{equation}

For the Casimir value $\omega = 3$, this reduces to
\begin{equation}
I_V^{(\omega=3)} = -\frac{16\pi r_0}{3\kappa},
\label{eq:IV_casimir}
\end{equation}
while for the Ellis--Bronnikov wormhole with $\omega = 1$, we obtain
\begin{equation}
I_V^{(\omega=1)} = -\frac{8\pi r_0}{\kappa}.
\label{eq:IV_ellis}
\end{equation}

Several features of the result~\eqref{eq:IV_result} merit discussion. First, unlike the original Morris--Thorne wormhole with $b(r) = r_0$ and $\phi(r) = 0$, for which the volume integral diverges~\cite{Garattini2019}, the Casimir--like wormholes require only a finite amount of exotic matter to sustain the geometry. For $\omega=3$, this finiteness was already noted by Garattini for the Planckian Casimir wormhole~\cite{Garattini2019}, and Eq.~\eqref{eq:IV_result} extends it to arbitrary $\omega$ and to throats of arbitrary size. This finiteness is a direct consequence of the rapid fall--off of the energy density as $\rho \propto r^{-4}$. Second, the total exotic matter content scales linearly with the throat radius $r_0$, implying that macroscopic wormholes, while requiring more exotic matter in absolute terms, do so in a controlled manner: doubling the throat size merely doubles the exotic matter requirement. Third, the prefactor $(1 + \omega)/\omega$ reveals how the exotic matter content depends on the equation of state parameter. For large $\omega$, this factor approaches unity, yielding $I_V \to -4\pi r_0/\kappa$, whereas as $\omega \to 0^+$, the exotic matter requirement diverges, indicating that softer equations of state demand increasingly large amounts of NEC--violating matter. Finally, the negative sign of $I_V$ confirms that the null energy condition is violated on average throughout the spacetime, as required for traversable wormhole geometries.

It is instructive to compare the exotic matter requirements for wormholes of different sizes. Restoring physical units, the volume integral~\eqref{eq:IV_casimir} for the Casimir value $\omega=3$ becomes
\begin{equation}
I_V^{(\omega=3)} = -\frac{2 c^4 r_0}{3 G}.
\end{equation}
This quantity has dimensions of energy, and its mass equivalent is $|I_V^{(\omega=3)}|/c^2=2c^2r_0/(3G)$. Table~\ref{tab:exotic_matter} presents this mass equivalent for throat radii spanning from Planckian to astrophysical scales.

\begin{table}[h]
\caption{Total exotic matter content $|I_V|$, expressed as its mass equivalent $|I_V|/c^2$, for wormholes with Casimir--type matter ($\omega = 3$) at different scales. The Planckian case corresponds to the Casimir wormhole, while the remaining entries represent Casimir--like wormholes with the amplitude treated as a free parameter. For comparison, the mass of the Earth is approximately $6 \times 10^{24}$ kg and the solar mass is approximately $2 \times 10^{30}$ kg.}
\label{tab:exotic_matter}
\begin{indented}
\item[]\begin{tabular}{lcc}
\mr
\textbf{Scale} & \textbf{$r_0$} & \textbf{$|I_V|/c^2$ (kg)} \\
\mr
Planckian & $10^{-35}$ m & $\sim 10^{-8}$ \\
\mr
Nuclear & $10^{-15}$ m & $\sim 10^{12}$ \\
\mr
Macroscopic & $1$ m & $\sim 10^{27}$ \\
\mr
Stellar & $10^{3}$ m & $\sim 10^{30}$ \\
\mr
Astrophysical & $10^{6}$ m & $\sim 10^{33}$ \\
\mr
Tidal bound (section~\ref{sec:tidal}) & $5.5\times10^{7}$ m & $\sim 5\times10^{34}$ \\
\mr
\end{tabular}
\end{indented}

\end{table}

The results in Table~\ref{tab:exotic_matter} reveal that while Planckian wormholes require negligible amounts of exotic matter, macroscopic and astrophysical wormholes demand quantities comparable to planetary or stellar masses. Nevertheless, the finiteness of these requirements, in contrast to the infinite exotic matter needed for simpler wormhole models, makes Casimir--like wormholes considerably less demanding, in terms of total exotic matter, than models for which the volume integral diverges.


\subsection{Tidal forces and traversability constraints}
\label{sec:tidal}

Throughout this paper, the term traversable refers to the geometric conditions of Morris and Thorne listed in section~\ref{subsec:MT} and summarized in Table~\ref{tab:matter_types}, namely the presence of a throat satisfying the flare--out condition and the absence of horizons. In this subsection and the next, we examine the additional and more restrictive conditions under which the wormhole could be crossed by a human traveler, which we refer to as human traversability~\cite{MaldacenaMilekhin2021}. Let us now focus on the human traversability conditions of the Casimir--like wormholes given by Eqs.~\eqref{eq:bwh} and~\eqref{eq:ephi}. We shall suppose that a traveler journeys radially through the wormhole, beginning at rest in a space station in the lower universe and ending at rest in a space station in the upper universe.

An important condition for human traversability is that the acceleration felt by the traveler should not exceed Earth's gravity $g_\oplus \approx 9.8\,\text{m/s}^2$. As shown in Refs.~\cite{Morris1988,Visser1995}, the acceleration felt by a radially moving traveler is given by
\begin{equation}
a = \pm \sqrt{1 - \frac{b(r)}{r}}\,e^{-\phi(r)}\frac{d}{dr}\left(\gamma e^{\phi(r)}\right)c^2,
\label{eq:acceleration}
\end{equation}
where $\gamma = (1 - v^2/c^2)^{-1/2}$ and $v(r)$ is the radial velocity of the traveler as measured by a static observer at $r$. For the Casimir--like wormhole~\eqref{eq:bwh}--\eqref{eq:ephi} this becomes
\begin{equation}
|a| = \sqrt{\frac{(r - r_0)(\omega r + r_0)}{\omega r^2}}\left|e^{-\phi}\frac{d(\gamma e^{\phi})}{dr}\right|c^2.
\label{eq:a_casimir}
\end{equation}

Since $\sqrt{1-b/r}\,d/dr=d/d\ell$, Eq.~\eqref{eq:acceleration} can be written as $a=\pm e^{-\phi}\,d(\gamma e^{\phi})/d\ell\;c^2$. The acceleration felt by the traveler is therefore, in general, finite and nonzero at the throat, and it vanishes identically only along a radial geodesic, for which $\gamma e^{\phi}$ is constant.

If the traveler maintains constant speed ($\gamma' = 0$), using $\phi'(r) = (\omega - 1)r_0/[2r(\omega r + r_0)]$, Eq.~\eqref{eq:a_casimir} reduces to
\begin{equation}
|a| = \frac{\gamma(\omega - 1)r_0}{2\sqrt{\omega}\,r^2}\sqrt{\frac{r - r_0}{\omega r + r_0}}\,c^2.
\label{eq:a_constant_v}
\end{equation}
For the Casimir value $\omega = 3$, this simplifies to
\begin{equation}
|a| = \frac{\gamma\,r_0}{\sqrt{3}\,r^2}\sqrt{\frac{r - r_0}{3r + r_0}}\,c^2.
\end{equation}
From this expression, one can show that the maximum acceleration occurs at
$r \approx 1.26\, r_0$, where the magnitude of the acceleration reaches
$|a|_{\max} \approx 0.085\, \gamma\, c^2 / r_0$. Requiring $|a|_{\max} \lesssim g_\oplus$ in the nonrelativistic limit ($\gamma \approx 1$) yields
\begin{equation}
r_0 \gtrsim \frac{0.085\,c^2}{g_\oplus} \approx 7.8 \times 10^{14}\,\text{m}.
\end{equation}
This constraint is extremely demanding. Its origin is physical. For $\omega=3$ the wormhole has a positive asymptotic mass, $M=b(\infty)c^2/(2G)=r_0c^2/(3G)$, about $2\times10^{4}$ solar masses for $r_0=10^{8}$~m, and a traveler moving slowly near the throat must counteract its gravitational pull, which for $\gamma\simeq1$ coincides with the acceleration of static observers, $\sqrt{1-b/r}\,\phi'c^2$. For $r_0=5.5\times10^{7}$~m this amounts to about $1.4\times10^{7}\,g_\oplus$, so that no slow traversal of such a wormhole is compatible with the acceleration constraint.

The natural way to satisfy this constraint is a radial free fall. Freely falling travelers along radial geodesics have also been used to assess the tidal traversability of other wormhole geometries, as in the case of NUT wormholes~\cite{Clement2016}. Along a radial geodesic, energy conservation gives $\gamma e^{\phi}=E$, so that Eq.~\eqref{eq:acceleration} yields $a=0$ identically, and the traveler only experiences tidal forces. For a traveler released from rest at a station located at $r=r_s$, one has $E=e^{\phi(r_s)}$, and the velocity profile is fixed by
\begin{equation}
\gamma(r)=e^{\phi(r_s)-\phi(r)}=e^{\phi(r_s)}\,\frac{3r+r_0}{3r}\qquad(\omega=3).
\label{eq:gamma_freefall}
\end{equation}
For stations located far from the throat, $e^{\phi(r_s)}\simeq1$, and the traveler reaches the throat with $\gamma\simeq4/3$, that is, $v\simeq0.66\,c$, independently of $r_0$. Since the geometry is symmetric with respect to the throat, the traveler then decelerates as it climbs out on the other side and comes to rest exactly at the radius $r_s$, where it can be retained by a second station. Gravity thus accelerates the traveler on the way in and decelerates it on the way out, and no rocket acceleration is required at any stage. The price is a relativistic crossing of the throat, which must be checked against the tidal constraints.

Another important condition for human traversability concerns the tidal accelerations felt by the traveler, which should not exceed Earth's gravitational acceleration. The radial and lateral tidal constraints are given by~\cite{Morris1988,Visser1995}
\begin{align}
&\left|\left(1 - \frac{b}{r}\right)\left[-\phi'' - (\phi')^2 + \frac{b'r - b}{2r(r-b)}\phi'\right]c^2\right||\xi| \lesssim g_\oplus, \label{eq:radial_tidal}\\[2mm]
&\left|\frac{\gamma^2 v^2}{2r^3}(b'r - b)\right||\xi| \lesssim g_\oplus, \label{eq:lateral_tidal}
\end{align}
where $\xi \approx 2\,\text{m}$ represents the size of the traveler's body. We note that for zero tidal--force wormholes, characterized by $\phi(r)=0$, the radial tidal acceleration vanishes identically.

For the Casimir--like wormholes, the radial tidal constraint~\eqref{eq:radial_tidal} reduces at the throat to
\begin{equation}
\frac{c^{2}\,|\omega-1|}{4|\omega|\,r_0^{2}}\,|\xi| \lesssim g_\oplus.
\label{eq:Rtrtr_throat}
\end{equation}
For $\omega=3$, the radial tidal acceleration along the trajectory is maximal at the throat. Note that for the Ellis--Bronnikov wormhole ($\omega = 1$), the radial tidal acceleration vanishes identically at the throat. Table~\ref{tab:tidal} presents the minimum throat radius allowed by this constraint for different values of $\omega$.

\begin{table}[h]
\caption{Minimum throat radius allowed by the radial tidal constraint at the throat ($\xi = 2\,\text{m}$, $g_\oplus = 9.8\,\text{m/s}^2$).}
\label{tab:tidal}
\begin{indented}
\item[]\begin{tabular}{cc}
\mr
$\omega$ & $r_0^{\min}$ \\
\mr
$1$ (Ellis--Bronnikov) & No constraint \\
\mr
$3$ (Casimir) & $5.5 \times 10^7\,\text{m}$ \\
\mr
$-3/2$ (Phantom) & $8.7 \times 10^7\,\text{m}$ \\
\mr
\end{tabular}
\end{indented}

\end{table}

Evaluating Eq.~\eqref{eq:lateral_tidal} at the throat, we obtain the corresponding lateral tidal constraint, 
\begin{equation}
\frac{\gamma^2 v^2 |1 + \omega|}{2|\omega|\,r_0^2}\,|\xi| \lesssim g_\oplus.
\label{eq:lateral_throat}
\end{equation}
Solving for the traveler's maximum speed, we obtain
\begin{equation}
v \lesssim \frac{r_0 K}{\sqrt{1 + (r_0 K/c)^2}},
\label{eq:v_exact}
\end{equation}
where
\begin{equation}
K = \sqrt{\frac{2|\omega|\,g_\oplus}{|1 + \omega|\,\xi}}.
\label{eq:K_def}
\end{equation}
For the Casimir value $\omega = 3$ with $\xi = 2\,\text{m}$, we have $K \approx 2.71\,\text{s}^{-1}$. Considering a tidal--safe wormhole with $r_0 = 5.5 \times 10^7\,\text{m}$, we find $r_0 K/c \approx 0.50$, which gives $v \lesssim 1.3 \times 10^8\,\text{m/s} \approx 0.45\,c$, with a corresponding Lorentz factor $\gamma \approx 1.12$.

For the free--fall traversal described above, however, the speed at the throat is not a free parameter but is fixed by Eq.~\eqref{eq:gamma_freefall}, $\gamma\simeq4/3$, so that $\gamma^2v^2=(\gamma^2-1)c^2\simeq7c^2/9$, which exceeds the bound obtained for $r_0=5.5\times10^{7}$~m. The lateral constraint~\eqref{eq:lateral_throat} then requires $r_0\gtrsim9.8\times10^{7}$~m. Since the lateral tidal acceleration decreases away from the throat along the trajectory, this is the relevant bound. For a freely falling traveler, the minimum throat radius compatible with all the constraints considered here is therefore $r_0\simeq10^{8}$~m, set by the lateral rather than the radial tidal acceleration.

\subsection{Transit time}
\label{sec:transit}

We now analyze the total travel time between station~1 and station~2. The proper radial distance from the throat to a coordinate radius $r$ is given by Eq.~\eqref{eq:proper_distance}. For the Casimir--like wormhole described by the shape function~\eqref{eq:bwh}, this distance takes the form
\begin{equation}
\ell(r) = \pm \int_{r_0}^{r} 
\sqrt{\frac{\omega\, r'^2}{(r' - r_0)(\omega r' + r_0)}}\, dr'.
\label{eq:ell_integral}
\end{equation}

For the free--fall traversal described in section~\ref{sec:tidal}, the location of the stations requires some care. The standard Morris--Thorne criterion, $1-b(r)/r\simeq0.99$, places the stations at $r\approx67\,r_0$ for $\omega=3$, but, owing to the positive mass of the wormhole, observers at rest at that radius would experience a gravitational acceleration of about $7\times10^{3}\,g_\oplus$ for $r_0=10^{8}$~m. We therefore require instead that observers at rest at the stations experience at most $g_\oplus$, that is, $\sqrt{1-b(r_s)/r_s}\,\phi'(r_s)\,c^2\le g_\oplus$. For $r_0=10^{8}$~m, this places the stations at $r_s\approx5.5\times10^{11}$~m, about $3.7$ astronomical units from the throat.

The time elapsed during the journey, as measured by the traveler and by the station clocks, is
\begin{equation}
\Delta\tau=2\int_{r_0}^{r_s}\frac{d\ell}{\gamma v},\qquad \Delta t_{\rm st}=2\int_{r_0}^{r_s}\frac{\gamma\,d\ell}{v},
\label{eq:transit_freefall}
\end{equation}
where $d\ell$ is the integrand of Eq.~\eqref{eq:ell_integral}, $\gamma(r)$ is given by Eq.~\eqref{eq:gamma_freefall}, and the factor $\gamma=e^{\phi(r_s)-\phi(r)}$ in $\Delta t_{\rm st}$ converts the local time of static observers into station time. Numerical integration gives $\Delta\tau\simeq\Delta t_{\rm st}\simeq6.1$~days for $r_0=10^{8}$~m. The two times nearly coincide because most of the journey takes place far from the throat, where the motion is nonrelativistic, and indeed the result agrees with the Newtonian free--fall estimate $2\times(\pi/2)\sqrt{r_s^3/(2GM)}$. Along this trajectory the felt acceleration vanishes identically, while the radial and lateral tidal accelerations, both maximal at the throat, are about $0.31\,g_\oplus$ and $0.95\,g_\oplus$, respectively. All the human traversability constraints considered here are therefore satisfied simultaneously for $r_0\gtrsim10^{8}$~m.

\subsection{Embeddings of equatorial slices of the Casimir--like wormholes}

The Casimir--like wormhole solutions are characterized by the shape function~\eqref{eq:bwh} and the redshift function~\eqref{eq:ephi}. The corresponding metric is given by
\begin{equation}
\begin{split}
ds^{2} &= -\left(\frac{\omega r}{\omega r + r_{0}}\right)^{\omega - 1}dt^{2} + \frac{dr^{2}}{1 - \dfrac{r_{0} - \dfrac{r_{0} (r - r_{0})}{\omega r}}{r}} \\
&\quad + r^{2} \left(d\theta^{2} + \sin^{2}\theta \, d\varphi^{2}\right),
\end{split}
\label{eq:mt_metric_b_phi}
\end{equation}
where $r_0$ denotes the throat radius and $\omega$ is a dimensionless parameter that characterizes both the wormhole geometry and the properties of the matter supporting the wormhole configuration.

We now consider the embedding of equatorial slices $t=\text{const}$ and $\theta=\pi/2$ of this metric in a Euclidean three--dimensional space. We employ the standard procedure that introduces the lift function $z(r)$ given by~\cite{Morris1988,Cataldo2025}
\begin{equation}
\frac{dz}{dr}=\pm \left(\frac{1}{r/b(r)-1} \right)^{1/2}, 
\label{z(r)}
\end{equation}
where the function $z(r)$ is also known in the literature as the embedding function and plays a fundamental role in the geometric visualization of wormholes. For the metric~\eqref{eq:mt_metric_b_phi}, Eq.~\eqref{z(r)} takes the following form:
\begin{equation}
\frac{dz(r)}{dr} = \pm \sqrt{\frac{(r_{0} + r(\omega - 1))r_{0}}{(r - r_{0})(\omega r + r_{0})}}.
\label{eq:embedding_casimir}
\end{equation}

From this expression, it is clear that at the throat, the derivative diverges as expected for wormhole geometries. For the embedding to exist, the expression under the square root must remain non--negative. This requirement leads to three distinct geometric regimes depending on the value of $\omega$, as summarized in Table~\ref{tab:embedding_regimes}.

\begin{table}[h]
\caption{Embedding regimes for Casimir--like wormholes as a function of the parameter $\omega$. The table shows the radial extent of the embedding and representative values used in the figures.}
\label{tab:embedding_regimes}

\begin{indented}
\item[]\begin{tabular}{cccc}
\mr\mr
Regime & $\omega$ range & Radial extent & Examples \\
\mr
I & $\omega \geq 1$ & $r_0 \leq r < \infty$ & $\omega = 1, 3$ \\
II & $\omega < -1$ & $r_0 \leq r < \infty$ & $\omega = -3/2$ \\
III & $0 < \omega < 1$ & $r_0 \leq r \leq r_{\max} = \frac{r_0}{1-\omega}$ & $\omega = 0.7, 0.8, 0.9$ \\
\mr\mr
\end{tabular}
\end{indented}
\end{table}

\begin{figure}
\includegraphics[scale=0.29]{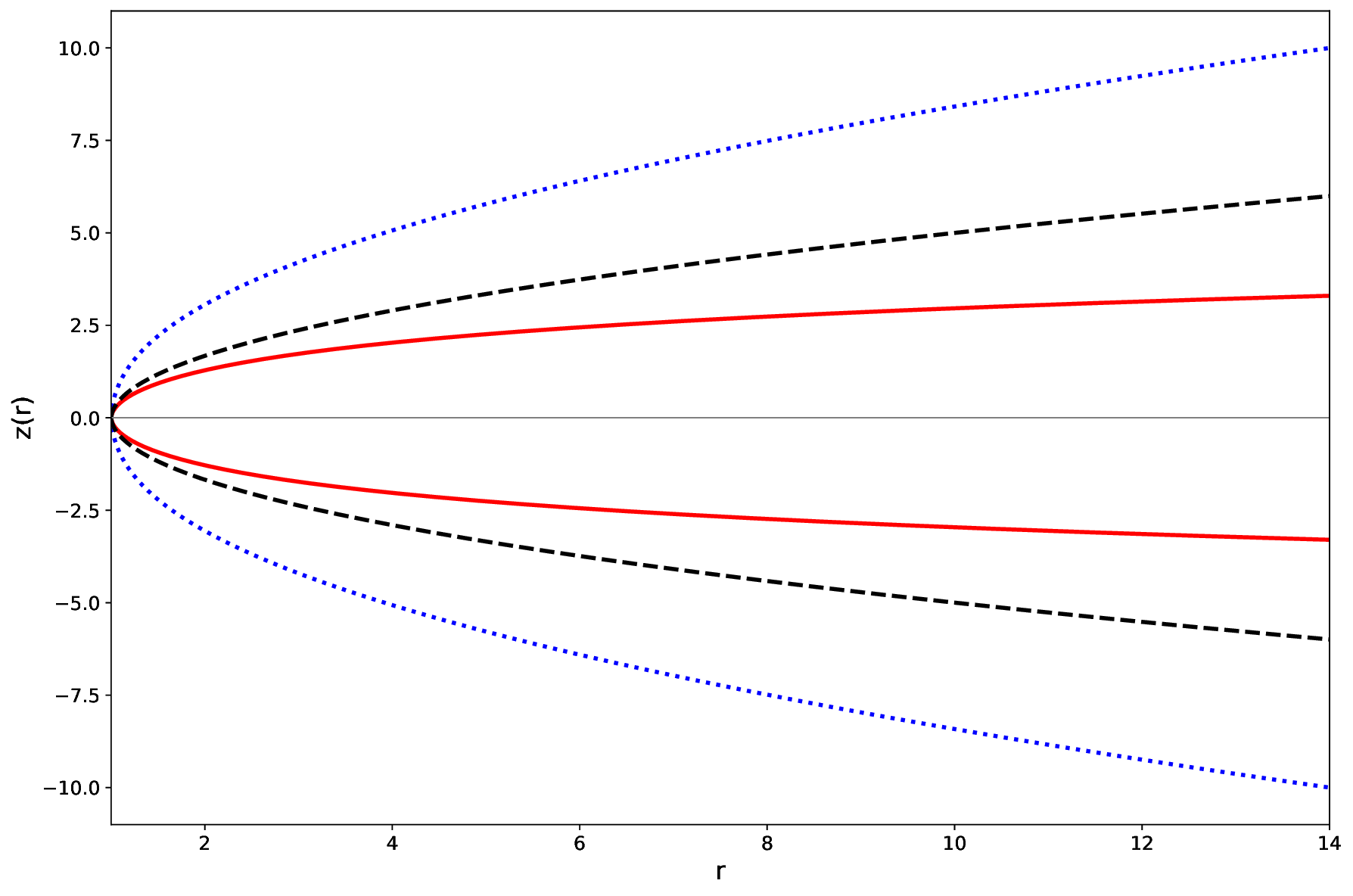}
                                                                                                                                                                                                                                                                                                                                                                                                                                                                                                                 \caption{Embedding diagrams for Casimir--like wormholes in Regimes I and II. The curves correspond to $\omega = -3/2$ (dotted line), $\omega = 1$ (solid line), and $\omega = 3$ (dashed line). For these configurations, the embedding extends from the throat at $r = r_0$ to spatial infinity, with the throat located at $z = 0$.}
\label{Fig15}
\end{figure}

For Regimes I ($\omega \geq 1$) and II ($\omega < -1$), the radicand in Eq.~\eqref{eq:embedding_casimir} remains positive for all $r > r_0$, ensuring that the embedding extends from the throat at $r = r_0$ to spatial infinity. This scenario is illustrated in Fig.~\ref{Fig15}, which shows the embedding diagrams for $\omega = -3/2$ (dotted line), $\omega = 1$ (solid line), and $\omega = 3$ (dashed line). These configurations represent wormholes that asymptotically approach flat space at large radii.

\begin{figure}
\includegraphics[scale=0.291]{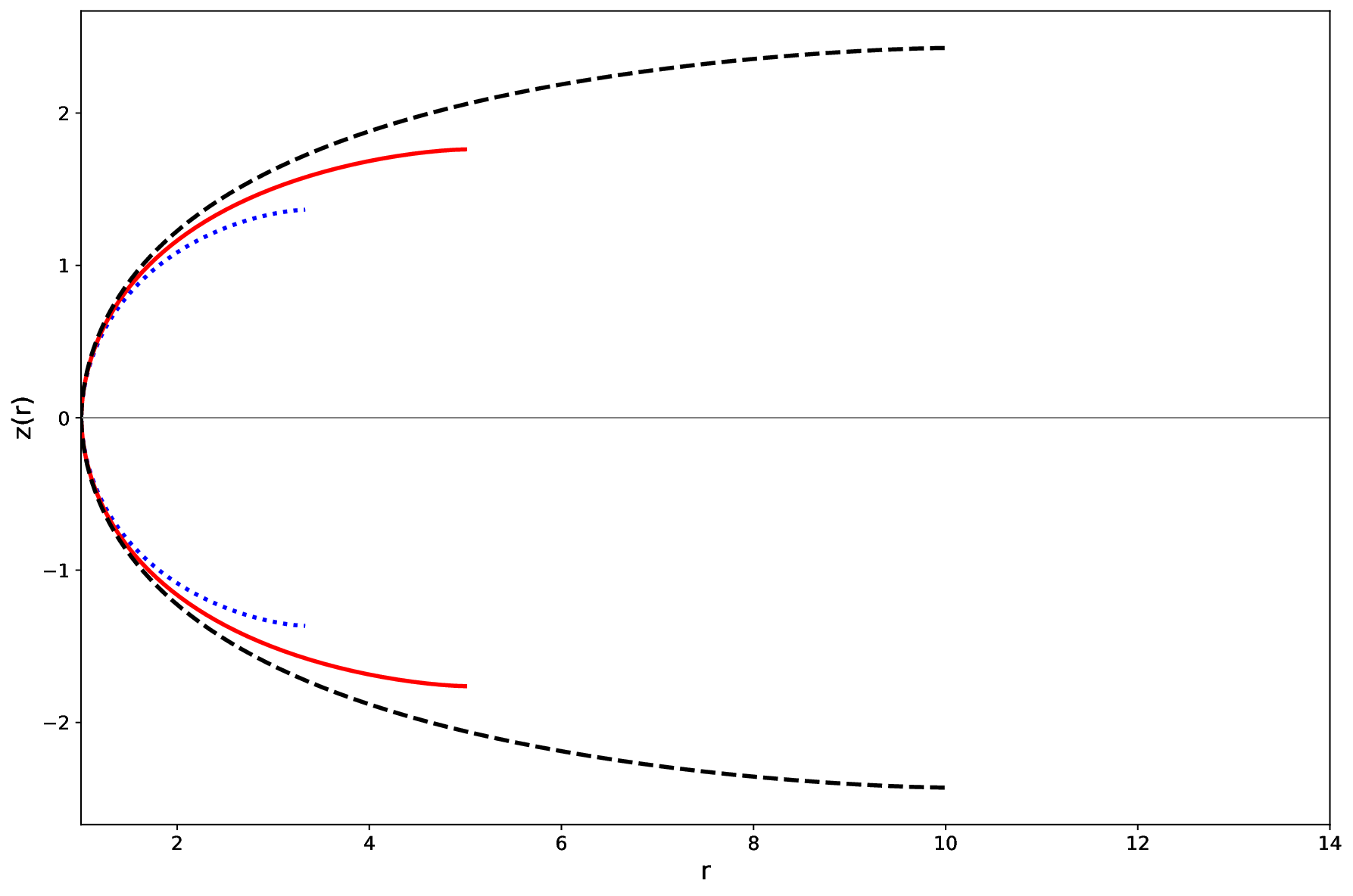}
\caption{Embedding diagrams for Casimir--like wormholes in Regime III with $0 < \omega < 1$. The curves correspond to $\omega = 0.7$ (dotted line), $\omega = 0.8$ (solid line), and $\omega = 0.9$ (dashed line). The embeddings extend from the throat at $r = r_0$ to maximum radii $r_{\max} = 10r_0/3$, $r_{\max} = 5r_0$, and $r_{\max} = 10r_0$, respectively, where the embedding function terminates.}
\label{Fig25} 
\end{figure}

\begin{figure}
\includegraphics[scale=0.35]{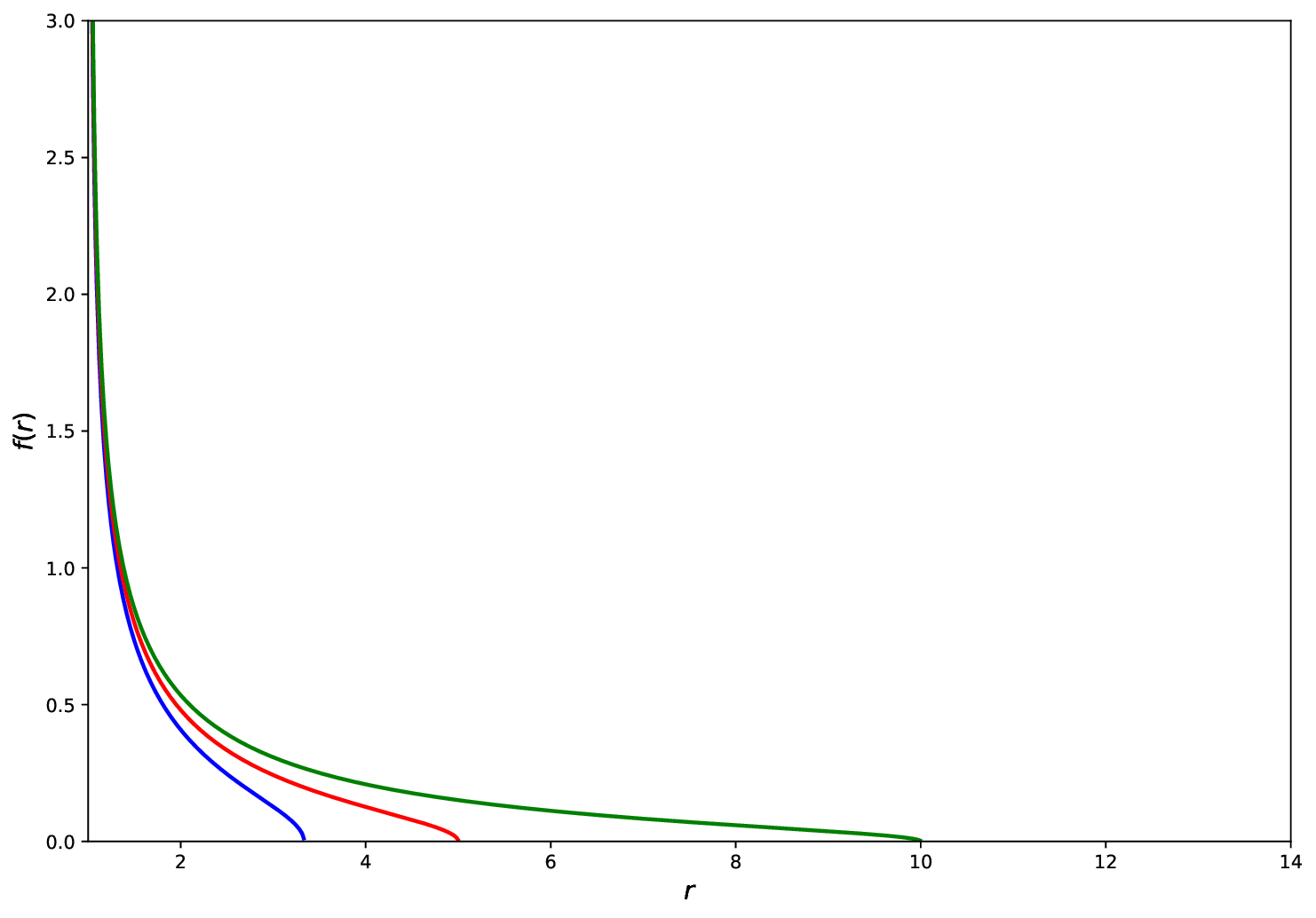}
\caption{Radicand $f(r) = \frac{(r_{0} + r(\omega-1))r_{0}}{(r-r_{0})(\omega r + r_{0})}$ from Eq.~\eqref{eq:embedding_casimir} as a function of $r$ for $\omega = 0.7$, $\omega = 0.8$ and $\omega = 0.9$ with $r_0 = 1$. The radicand vanishes at $r = 10/3$, $5$, and $10$, respectively, marking the maximum radial extent of the embeddings shown in Fig.~\ref{Fig25}.}
\label{Fig35}
\end{figure}

In contrast, for Regime III ($0 < \omega < 1$), the radicand becomes negative beyond a critical radius $r_{\max} = r_0/(1-\omega)$, limiting the embedding to a finite radial range $r_0 \leq r \leq r_{\max}$. Fig.~\ref{Fig25} displays the embedding diagrams for $\omega = 0.7$ (dotted line), $\omega = 0.8$ (solid line), and $\omega = 0.9$ (dashed line). For these parameter values, the maximum radii are $r_{\max} = 10r_0/3$ ($\omega = 0.7$), $r_{\max} = 5r_0$ ($\omega = 0.8$), and $r_{\max} = 10r_0$ ($\omega = 0.9$), respectively. This behavior is further illustrated in Fig.~\ref{Fig35}, which shows the radicand $f(r)$ of Eq.~\eqref{eq:embedding_casimir} as a function of $r$ for the same values of $\omega$ with $r_0 = 1$. As can be observed, the radicand vanishes precisely at $r = 10/3$, $5$, and $10$ for $\omega = 0.7$, $0.8$, and $0.9$, respectively, confirming that the embeddings terminate at these critical radii. 

For the cases shown in Fig.~\ref{Fig15}, the radicand $f(r)$ in Eq.~\eqref{eq:embedding_casimir} remains positive for all $r \geq r_0$, with $f(r) \rightarrow 0$ as $r \rightarrow \infty$.

\section{Conclusions}\label{sec:conclusions}
In this work, we have presented a comprehensive analysis of spherically symmetric spacetimes sourced by matter obeying a Casimir--type equation of state $p_r = \omega \rho$, with an energy density scaling as $\rho \propto r^{-4}$. In contrast to Garattini’s original approach, which is intrinsically tied to the quantum Casimir effect and its associated Planckian scales, we have adopted a phenomenological perspective in which the Casimir--type equation of state is treated as an effective description that may arise from yet unknown physics at macroscopic scales. This viewpoint is consistent with the standard methodology in modern wormhole physics, where one postulates the existence of exotic matter satisfying specific equations of state and studies the resulting geometric and traversability properties without committing to a particular microscopic origin. Within this framework, we have solved Einstein’s field equations without imposing traversability conditions a priori, obtaining the most general solution for this class of matter. The resulting spacetime geometries form a rich family that includes black holes, naked singularities, and traversable wormholes, parametrized by the integration constants and the equation of state parameter $\omega$.

A central finding of this work is that the complete algebraic structure of the Casimir stress--energy tensor, characterized by $\rho < 0$, $p_r = 3\rho$, and $p_t = -\rho$, emerges dynamically in the vicinity of the central naked singularity at $r = 0$. This result is particularly noteworthy because the relation $p_t = -\rho$ is not imposed as an input condition but arises directly from the Einstein field equations, and, for a macroscopic amplitude $A$, it is already approximately attained at radii where classical general relativity remains reliable. To our knowledge, this feature has not been previously recognized in the literature on Casimir spherically symmetric configurations.

Regarding the classification of matter sources capable of supporting traversable wormholes, we have established that only two classes satisfy the required flare--out condition: Casimir--type matter with $\omega > 0$ and phantom matter with $\omega < -1$. Quintessence and conventional dark energy are excluded because they fail to satisfy the strict inequality $b'(r_0) < 1$ at the throat. This classification clarifies which equations of state are compatible with traversable wormhole geometries within the $\rho \propto r^{-4}$ framework.

When the amplitude of the energy density is identified with the quantum Casimir energy scale, the traversability condition uniquely fixes the throat radius at $r_0 \approx 1.02 \, \ell_P$ for the  Casimir value $\omega = 3$. This result recovers Garattini's original finding~\cite{Garattini2019} as a particular case within our general framework, confirming that wormholes sourced by the quantum vacuum Casimir effect are necessarily of Planckian size. However, by treating the amplitude as a free parameter unconstrained by quantum considerations, we have demonstrated that the mathematical structure of Einstein's equations admits traversable wormhole solutions with throat radii ranging from microscopic to astrophysical scales. The energy density at the throat scales as $|\rho(r_0)| \propto r_0^{-2}$, implying that larger wormholes require progressively smaller energy densities to sustain the geometry.

A notable distinction between our approach and that of Garattini concerns the determination of the throat radius. In Garattini's treatment, as well as in subsequent generalizations, the throat radius is effectively fixed by imposing specific conditions on the equation of state parameter at the throat, which then constrains $r_0$ through consistency with the Casimir energy scale. This procedure constitutes a form of fine--tuning that ties the geometry to particular quantum scales. In contrast, our analysis treats $r_0$ as a free parameter from the outset, with the amplitude of the energy density adjusting accordingly through the traversability condition $A = -r_0^2/\omega$. This freedom allows us to explore the full landscape of Casimir--like wormhole solutions without restricting ourselves to Planckian regimes. We note that the traversable geometry itself coincides with the solution obtained by Garattini for a generic $\omega$~\cite{Garattini2019}. The new ingredients of the present analysis are the general solution obtained without imposing a throat, which reveals the Casimir--type structure near the origin, the treatment of the amplitude $A$ as a free parameter of either sign, which allows macroscopic throats and the phantom branch $\omega<-1$, and the analysis of the exotic matter content for arbitrary $\omega$, of human traversability and of physical realizability at macroscopic scales.

We have also shown that, unlike simpler wormhole models for which the volume integral quantifier diverges, the Casimir--like wormholes studied here require only a finite amount of exotic matter to sustain the geometry, extending to arbitrary $\omega$ and throat size a result obtained by Garattini for the Planckian case $\omega=3$~\cite{Garattini2019}. The total exotic matter content scales linearly with the throat radius, $I_V \propto r_0$, a consequence of the rapid fall--off of the energy density as $\rho \propto r^{-4}$. This finiteness makes Casimir--like wormholes considerably less demanding, in terms of total exotic matter, than models requiring infinite amounts of NEC--violating matter, although for macroscopic throats the required amount remains of planetary to stellar order.

We have also analyzed the conditions for human traversability. For the Casimir value $\omega=3$, the wormhole has a positive asymptotic mass, so that a slow traversal near the throat would require accelerations far beyond $g_\oplus$. A radial free fall avoids this problem, since the traveler experiences no acceleration at all. Released from rest at a station, the traveler reaches the throat with $v\simeq0.66\,c$ and arrives at rest at the symmetric station on the other side. The lateral tidal constraint then requires $r_0\gtrsim10^{8}$~m, in which case the journey takes about six days, with the stations located a few astronomical units from the throat.

Finally, we have identified three distinct geometric regimes for the embedding of equatorial slices depending on the value of $\omega$: for $\omega \geq 1$ and $\omega < -1$, the embedding extends from the throat to spatial infinity, while for $0 < \omega < 1$, the embedding is limited to a finite radial range. These regimes provide a geometric visualization of how the equation of state parameter influences the global structure of the wormhole spacetime.

We emphasize that the present analysis is purely geometric and classical. We have not carried out a dynamical or perturbative stability analysis of the wormhole throat. As discussed in section~\ref{subsec:eos}, the radial relation $dp_r/d\rho=\omega=3$ trivially exceeds unity, being a direct restatement of the imposed equation of state rather than a property derived from an identified dynamical field; since we do not specify which matter realizes this equation of state, this formal value cannot by itself be taken as a demonstration of superluminal propagation, as illustrated by the genuine Casimir effect itself, whose photon degrees of freedom propagate at the speed of light despite sharing the same $\omega=3$. Establishing the actual causal behavior of these configurations rigorously would require either a specific field--theoretic realization of the anisotropic source or a dynamical/perturbative treatment of the throat, both of which we leave for future work.

In summary, we have demonstrated that the space of Casimir--like wormhole solutions is considerably richer than previously recognized. The physical realizability of macroscopic Casimir--like wormholes remains an open question, intimately connected to the broader problem of exotic matter sources in general relativity. Nevertheless, the existence of these solutions within the classical theory suggests that the search for traversable wormholes need not be confined to Planckian scales, opening intriguing possibilities for future theoretical and observational investigations.

\section*{Acknowledgments}
This work was supported by the Dirección de Investigación y Creación Artística at the Universidad del Bío-Bío through grants No. RE2320220 and GI2310339.

{\bf Data Availability Statement:} This work is purely theoretical and does not involve the generation or analysis of experimental or observational data.

\end{document}